\documentclass[aps,prc,10pt,amsmath,amssymb,twocolumn,floatfix,nofootinbib,superscriptaddress,hyperref]{revtex4-1}

\usepackage{graphicx}
\usepackage{dcolumn}
\usepackage{bm}
\usepackage{hyperref}
\usepackage{lipsum}
\usepackage{todonotes}
\usepackage{xcolor}
\usepackage{textcomp}
\allowdisplaybreaks

\newcommand{\nj}[9]{ \begin{Bmatrix}
  #1 & #2 & #3 \\
  #4 & #5 & #6 \\
  #7 & #8 & #9
\end{Bmatrix} }

\newcommand{\sj}[6]{ \begin{Bmatrix}
  #1 & #2 & #3 \\
  #4 & #5 & #6
\end{Bmatrix} }

\newcommand{\cg}[6]{C^{#1 #2 #3}_{#4 #5 #6}}

\newcommand{\opx}{X}
\newcommand{\opy}{Y}
\newcommand{\opz}{Z}

\newcommand{\fm}{\ensuremath{\textrm{fm}}}
\newcommand{\fmi}{\ensuremath{\textrm{fm}^{-1}}}

\newcommand{\unnorm}[1]{\breve{#1}}

\begin{document}

\preprint{APS/123-QED}

\title{\normalsize Ab initio electromagnetic observables with the in-medium similarity renormalization group}

\author{N.~M.~Parzuchowski}
 \email{parzuchowski.1@osu.edu}
 \affiliation{Department of Physics, The Ohio State University, Columbus, OH 43210, USA} 
 \affiliation{Facility for Rare Isotope Beams and Department of Physics and Astronomy, Michigan State University, East Lansing, MI 48844, USA }%
\author{S.~R.~Stroberg}%
 \email{sstroberg@triumf.ca}
 \affiliation{TRIUMF 4004 Wesbrook Mall, Vancouver BC V6T 2A3 Canada}%
 \affiliation{Physics Department, Reed College, Portland OR, 97202, USA}
\author{P.~Navr\'atil}%
 \email{navratil@triumf.ca}
\affiliation{TRIUMF 4004 Wesbrook Mall, Vancouver BC V6T 2A3 Canada}%

\author{H.~Hergert}%
 \email{hergert@nscl.msu.edu}
\affiliation{Facility for Rare Isotope Beams and Department of Physics and Astronomy, Michigan State University, East Lansing, MI 48844, USA }%

 \author{S.~K.~Bogner}%
\email{bogner@nscl.msu.edu}
\affiliation{Facility for Rare Isotope Beams and Department of Physics and Astronomy, Michigan State University, East Lansing, MI 48844, USA }%

\date{\today}

\begin{abstract}
We present the formalism for consistently transforming transition operators within the in-medium similarity renormalization group framework.
We implement the operator transformation in both the equations-of-motion and valence-space variants, and present first results for electromagnetic transitions and moments in medium-mass nuclei using consistently-evolved operators, including the induced two-body parts. These results are compared to experimental values, and---where possible---the results of no-core shell model calculations using the same input chiral interaction.
We find good agreement between the equations-of-motion and valence space approaches.
Magnetic dipole observables are generally in reasonable agreement with experiment, while the more collective electric quadrupole and octupole observables are significantly underpredicted, often by over an order of magnitude, indicating missing physics at the present level of truncation.

\end{abstract}

\pacs{Valid PACS appear here}
\maketitle

\section{\label{sec:Introduction}Introduction}
Understanding the observed properties of atomic nuclei based upon the underlying hadronic degrees of freedom has long been a major goal of nuclear structure theory. Achieving this goal has become especially important as nuclei become laboratories in the search for physics beyond the Standard Model~\cite{Engel1992,Avignone2008,Menendez2012a,Gando2016,Engel0vBBreview}.
In the treatment of the nuclear physics relevant for these searches, the more traditional phenomenological approaches to nuclear physics---despite their tremendous success in predicting and interpreting existing nuclear data \cite{Brown2006a,Caurier2005}---suffer from a lack of guidance as to how to incorporate new physics and make meaningful predictions.
This is largely due to the fact that, by definition, there is no data for these processes upon which to fix phenomenological parameters.
One promising path forward is to construct nuclei {\it ab initio}, starting from the underlying degrees of freedom rooted in the Standard Model.
The two main tasks in this approach are the formulation of appropriate interactions between nucleons, and the solution of the resulting many-body problem with sufficient accuracy. Substantial progress has been made on the former difficulty by the application of chiral effective field theory (EFT) \cite{Epelbaum:2009ve,Machleidt:2011bh,Epelbaum2015,Entem:2015qf,Entem:2015pw}, though much work certainly remains.

On the many-body front, methods such as the no-core shell model (NCSM)~\cite{Navratil2007b,Navratil2009,Barrett2013} and quantum Monte Carlo (QMC)~\cite{Carlson2015} provide exact solutions for $p$-shell nuclei up to finite basis effects and sampling errors. While the application of renormalization group ideas~\cite{Bogner2007,Jurgenson2009,Bogner:2010pq,Roth2014b} has helped extend the reach of the NCSM, both of these methods encounter prohibitive computational scaling for medium-mass nuclei.

Another class of approximate but systematically improvable many-body methods, namely coupled-cluster (CC)~\cite{Hagen2014,Jansen2011,Binder2013}, self-consistent Green's functions (SCGF)~\cite{Cipollone2013,Soma2014,Cipollone2015,Soma2014a}, many-body perturbation theory (MBPT)~\cite{Hjorth-Jensen1995,Tsunoda2014,Simonis2016,Tichai2016}, and the in-medium similarity renormalization group (IMSRG)~\cite{Tsukiyama2011,Hergert2016,Bogner2014,Hergert:2016me}, have enabled applications to nuclei beyond the $fp$-shell~\cite{Soma2013,Binder2014,Hergert2014,Hagen2016b,Tichai2016,Simonis2017}.
Each of these methods may be formulated in terms of summed Goldstone diagrams (including some classes of diagrams to all orders), and each employs normal ordering with respect to a reference state in order to approximately treat three- and higher-body terms.
With these methods,
immense progress has been made in the calculation of nuclear binding energies, radii, and excited state spectra, where it is now possible to calculate these observable quantities consistently using two- and three-nucleon forces throughout the expanses of the medium-mass nuclear landscape.
At the present time, the deficiencies in the nuclear interactions have become the main source of error for many calculations, as opposed to truncation errors in the solution of the many-body problem.

As alluded to above, a major advantage of {\it ab initio} methods which start from chiral EFT is the possibility to obtain transition operators consistent with a given interaction.
A consistent treatment of operators is essential to address open questions in nuclear physics such as the source of axial-vector quenching in-medium~\cite{Wildenthal1983a,Martinez-Pinedo1996}
, and to do away with phenomenological concepts such as effective charges for $E2$ transitions.
It will also be indispensable for reliably calculating quantities relevant for searches for physics beyond the Standard Model, such as neutrinoless double beta decay~\cite{Engel0vBBreview}.
Finally, it remains to be demonstrated that the success of diagrammatic-expansion methods in calculating energies and radii carries over to other observables.

The effort to obtain consistent effective operators for use in nuclear structure calculations is certainly not new~ (see, e.g., \cite{DaProvidencia1964,Brandow1967,LoIudice1971,Tucson1975,Ellis1977}), and has long been a difficult problem for nuclear theory, though some progress has been made in recent years
~\cite{Anderson:2010br,Paar:2006zf,Schuster:2014rg,Stetcu:2005qh,Navratil1997,More2015}.
The IMSRG presents a straightforward framework for deriving consistent effective operators, because it is formulated in terms of a series of unitary transformations.
In order to reduce the storage needed for calculations, the IMSRG, like the other diagrammatic expansion methods, is generally formulated in an angular momentum coupled basis.
As a result, additional formal developments are required for the treatment of spherical tensor operators---i.e., operators that carry angular momentum---which are necessary for the calculation of transition strengths, electromagnetic moments and response functions.
In this work, we present a streamlined effective operator formalism for spherical tensors,  using the recently developed equations-of-motion IMSRG~\cite{Parzuchowski2017} (EOM-IMSRG) and valence-space IMSRG~\cite{Tsukiyama2012,Bogner2014,Stroberg2017} (VS-IMSRG).
The two methods offer complementary approaches to the problems of nuclear spectroscopy and decay, each with different benefits and drawbacks:
The EOM-IMSRG works with large single-particle spaces, but limits the type of particle-hole excitations, while VS-IMSRG treats \emph{all} particle-hole excitations in a small single-particle valence space exactly, but relies on a truncated IMSRG decoupling to account for excitations outside of that valence space. As we will discuss in the following, operators that appear in the IMSRG flow equations are truncated at the two-body level, and higher induced operators are neglected.
We will demonstrate that both methods are capable of consistently describing excited states and transitions for a certain class of states. In some cases we find results consistent with experiment, while in others we make note of discrepancies.

This work is organized as follows. In section~\ref{sec:Formalism}, we  give the relevant commutator expressions for  the calculation of effective tensor operators, and  lay out the formalism  for the EOM-IMSRG and VS-IMSRG. In section ~\ref{sec:Results}, we present results of calculations of transitions and moments for several nuclei ranging in mass from the deuteron to $^{60}$Ni, and we present conclusions in section~\ref{sec:Summary}. 

\section{\label{sec:Formalism}Formalism}
Here, we lay out the framework for evaluating matrix elements of spherical tensor operators in the IMSRG. For a review of the theory and formalism of IMSRG, we refer the reader to Ref.~\cite{Hergert2016}. 

\subsection{\label{subsec:Commutator_Formalism}Commutator expressions}
The main new development required for the transformation of tensor operators is the expression for the commutator between an operator of spherical tensor rank $\lambda$ with a scalar operator ($\lambda=0$).
We truncate all operators at the two-body level in the following discussion.
We write a scalar operator $\mathcal{S}$ in normal-ordered form as
\begin{equation}
\mathcal{S} = S_{0\textrm{b}}
+ \sum_{pq} S_{pq}\{a^{\dagger}_pa_q\}
+ \frac{1}{4} \sum_{pqrs} \unnorm{S}_{pqrs}^{J} \{a^{\dagger}_p a^{\dagger}_q a_s a_r \}.
\label{eq:define_S}
\end{equation}
The braces $\{\}$ indicate normal ordering with respect to the reference state $|\Phi_0\rangle$.
The zero-body term is given by $S_{0\textrm{b}}=\langle \Phi_0 |\mathcal{S}| \Phi_0\rangle$.
The coefficients $S_{pq}$ and $\unnorm{S}^{J}_{pqrs}$ are defined by 
\begin{gather}
S_{pq} \equiv \langle p | S | q \rangle \\
\unnorm{S}^{J}_{pqrs} \equiv \langle (pq) J | \unnorm{S} |(rs) J \rangle \,.
\end{gather}
Our two-body states are antisymmetrized but unnormalized, so that expressions may be written in terms of unrestricted sums. The unnormalized two-body matrix elements, indicated by a breve\footnote{In previous works, we have indicated unnormalized two-body matrix elements with a tilde ($\sim$). However, to avoid confusion in the present work we reserve the tilde to indicate spherical tensor annihilation operators.} \textasciibreve, are related to conventional normalized matrix elements via
\begin{equation}
\unnorm{S}^{J}_{pqrs} \equiv \sqrt{(1+\delta_{pq})(1+\delta_{rs})} S^{J}_{pqrs}
\end{equation}

We write a spherical tensor operator $\mathcal{T}^{\lambda}_{\mu}$ of rank $\lambda$ and projection $\mu$ as 
\begin{equation}
\begin{aligned}
\mathcal{T}^{\lambda}_{\mu} = T^{\lambda}_{0\textrm{b}} &+ \sum_{pq} T^{\lambda}_{pq}  \frac{[a^{\dagger}_p \times \tilde{a}_q]^{\lambda}_{\mu}}{\sqrt{2\lambda+1}}\\
&+ \frac{1}{4} \sum_{pqrs} \sum_{J_1 J_2} \unnorm{T}^{(J_1 J_2) \lambda}_{pqrs} \frac{\left[A^{\dagger J_1}_{pq}\times \tilde{A}^{J_2}_{rs}\right]^{\lambda}_{\mu}}{\sqrt{2\lambda+1}},
\label{eq:define_T}
\end{aligned}
\end{equation}
where $[\times]$ indicates a tensor product.
Note that a tensor operator ($\lambda \ne 0$) that is normal-ordered with respect to a spherical reference state (as  used in all calculations here) will have a zero-body piece $T^{\lambda}_{0\textrm{b}}=0$.
The tilde in eq.~(\ref{eq:define_T}) indicates the usual transformation of the annihilation operator $a_p$ to a spherical tensor operator~\cite{Bohr1969,Suhonen2007}:
\begin{equation}
\tilde{a}_p \equiv \tilde{a}_{(j_p,m_p)} = (-1)^{j_p+m_p}a_{(j_p,-m_p)}.
\end{equation}
$A^{\dagger JM}_{pq}$ is a creation operator for a two-particle state with total angular momentum $J$ and projection $M$:
\begin{equation}
\begin{aligned}
A^{\dagger JM}_{pq} |0\rangle &\equiv \left[a^{\dagger}_p \times a^{\dagger}_q \right]^{J}_{M}|0\rangle\\
&= | (pq)JM \rangle \\
\end{aligned}\,,
\end{equation}
with a corresponding definition for $\tilde{A}^{JM}_{rs}$
\begin{equation}
\tilde{A}^{JM}_{rs} = \left[\tilde{a}_s \times \tilde{a}_r\right]^{J}_{M}
= (-1)^{J-M}A^{J-M}_{rs}.
\end{equation}
The coefficients $T^\lambda_{pq}$ and $\unnorm{T}^{(J_1 J_2) \lambda}_{pqrs}$ are defined by the following reduced matrix elements, using the convention of Edmonds~\cite{Edmonds1960,Suhonen2007},
\begin{gather}
T^{\lambda}_{pq} \equiv \langle p \| T^{\lambda} \| q \rangle  \\
\unnorm{T}^{(J_1 J_2) \lambda}_{pqrs} \equiv \langle (pq)J_1 \| \unnorm{T}^{\lambda} \| (rs)J_2 \rangle.
\end{gather}
The commutator $\mathcal{C}^\lambda_\mu$ of the operators $\mathcal{S}$ and $\mathcal{T}^{\lambda}_{\mu}$ will be a spherical-tensor operator of rank $\lambda$:
\begin{equation}
\mathcal{C}^{\lambda}_{\mu} \equiv [\mathcal{S},\mathcal{T}^{\lambda}_{\mu}] = 
 \mathcal{S}\mathcal{T}^{\lambda}_{\mu} - \mathcal{T}^{\lambda}_{\mu}\mathcal{S}
 \label{eq:define_C}
\end{equation}
The coefficients $C^{\lambda}_{pq}$ and $\unnorm{C}^{(J_1J_2)\lambda}_{pqrs}$ are given by equations (\ref{eq:comm1body}) and (\ref{eq:comm2body}) in appendix~\ref{sec:APP_products}. 

\subsection{\label{subsec:EOM_Formalism}Equations-of-motion IMSRG}
In the equations-of-motion (EOM) formulation of the IMSRG, we first perform a single reference ground state calculation, which maps the reference $| \Phi_0 \rangle$ to the ground state $|\Psi_0\rangle$ via a continuous sequence of unitary transformations $U(s)$ that are labeled by the flow parameter $s$. We then describe the excited states in the IMSRG-transformed frame using a ladder operator $\bar{X}^{\dagger}_\nu$ acting on the reference state
\begin{equation}
U(\infty)|\Psi_\nu\rangle = \bar{X}_\nu^{\dagger}(J^\Pi)|\Phi_0 \rangle.
\end{equation}
Here the bar indicates that the ladder operator is expressed in the transformed frame.
The Schr\"odinger equation for the IMSRG rotated Hamiltonian $\bar{H}$ may then be written as
\begin{equation}\label{eq:EOM_eigval}
[\bar{H},\bar{X}^{\dagger}_{\nu}(J^\Pi)] |\Phi_0 \rangle = (E_\nu - E_0)\bar{X}^{\dagger}_\nu(J^\Pi)|\Phi_0 \rangle.
\end{equation}
As a result of the ground-state decoupling, there is no correlation between the ground state and excited states in the rotated frame, so $\bar{X}^\dagger_{\nu}(J^\Pi)$ will consist only of excitation operators of the form $a^\dagger_{a}a^\dagger_{b} \cdots a_{i} a_{j} \cdots$, where $a,b,c,\ldots$ and $i,j,k,\ldots$ denote orbitals that are unoccupied and occupied, respectively, in the reference state.
Note that evaluating the l.h.s. of (\ref{eq:EOM_eigval}) requires a scalar-tensor commutator as defined in (\ref{eq:define_C}).

Calculations of this type are subject to two sources of systematically improvable error, namely truncations of the IMSRG equations and truncations of the EOM ladder operator. In this work, both truncations will be made at the two-body level (EOM(2)-IMSRG(2) $\equiv$ EOM-IMSRG(2,2)). The normal ordering with respect to the reference state is crucially important to control the quality of these truncations, because it allows us to retain in-medium contributions from 3N forces in the normal ordered zero-, one-, and two- body pieces of our operators. Beyond the IMSRG framework, the truncation of input interactions and operators at the normal-ordered two-body level is known as the normal-ordered two-body (NO2B) approximation \cite{Hagen:2007zc,Roth:2012qf,Binder2014,Ekstrom2014a}. 
	
Our ladder operators are linear combinations of one- and two-body excitation operators coupled to desired spin $J^\Pi$

\begin{equation}
\begin{aligned}
\bar{X}^\dagger_\nu(J^\Pi M) &= \sum_{ai} X^J_{ai}(\nu)  \frac{[a^{\dagger}_a \times \tilde{a}_i]^{J}_{M}}{\sqrt{2 J +1}} \\
+ \frac{1}{4} &\sum_{abij} \sum_{J_1 J_2} \unnorm{X}^{(J_{1}J_{2})J}_{abij}(\nu) \frac{\left[A^{\dagger J_{2}}_{ab}\times \tilde{A}^{J_{1}}_{ij}\right]^{J}_{M}}{\sqrt{2 J +1}}.
\label{eq:define_X}
\end{aligned}
\end{equation}
The amplitudes $X^J_{ai}(\nu) $ and $\unnorm{X}^{(J_{1}J_{2})J}_{abij}(\nu)$, as well as excitation energies, are obtained by solving the eigenvalue problem (\ref{eq:EOM_eigval}).
Note that this formulation is equivalent to configuration interaction with singles and doubles (CISD), i.e. diagonalizing the transformed Hamiltonian in the space of 1p1h and 2p2h excitations out of $|\Phi_0\rangle$.

To quantify the importance of the EOM ladder operator truncation, we compute the 1p1h partial norms,
\begin{equation}\label{eq:EOM_1p1h_norm}
n_\nu(1p1h) = \sqrt{\sum_{ia} |\bar{X}^J_{ai}(\nu)|^2}. 
\end{equation}
For states with $n_\nu(1p1h)$ near one, we expect small error in the EOM portion of the calculation. A small 1p1h partial norm indicates that the rotated wave-function for the state in question contains higher-order many-body excitations which are not captured by the ladder operator in~(\ref{eq:define_X}). 

Operator matrix elements for transitions to the ground state may be written 
\begin{equation}
\begin{aligned}
M_{0\nu} &= \langle \Phi_0 || [\bar{O}^\lambda \times \bar{X}^\dagger_\nu(J^\Pi_\nu)]^0|| \Phi_0 \rangle   \\  
&= \delta_{\lambda J_\nu} (-1)^{J_\nu}\Biggr[\sum_{ai} \frac{X_{ai}(\nu,J_\nu^\Pi) }{\sqrt{2J + 1}} O_{ai}(\lambda,\Pi)\\ 
&+ \frac{1}{4}\sum_{abij}\sum_{J_1 J_2} \frac{\unnorm{X}^{J_1 J_2}_{abij}(\nu,J_\nu^\Pi) }{\sqrt{2J + 1}}   \unnorm{O}^{J_1 J_2}_{abij}(\lambda,\Pi)\Biggr]\,, 
\label{eq:ME_EOM_ground}
\end{aligned}
\end{equation}
and for transitions between excited states, or expectation values of excited states, 
\begin{equation}
M_{\mu \nu} = \langle \Phi_0 ||[ \bar{X}_\mu(J^\Pi_\mu) \times [\bar{O}^\lambda \times \bar{X}^\dagger_\nu(J^\Pi_\nu)]^{J_\mu}]^0|| \Phi_0 \rangle\,.     
\label{eq:ME_EOM_general}
\end{equation}
Equation~(\ref{eq:ME_EOM_general}) requires the calculation of the full tensor product 
\begin{equation}
\mathcal{Y}^J_M \equiv [\bar{\mathcal{O}}^{\lambda} \times \bar{X}^\dagger_\nu(J_\nu)]^J_M = \sum_{M_\nu \mu} C^{\lambda J_\nu J}_{\mu M_\nu M} \bar{\mathcal{O}}^\lambda_\mu \bar{X}^\dagger_\nu( J_\nu M_{\nu}).  
\label{eq:tensor_prod_def}
\end{equation}
The matrix elements of $\mathcal{Y}$ are given by equations~\ref{eq:tensor_prod_1b} and~\ref{eq:tensor_prod_2b} in appendix~\ref{sec:APP_products}.
In equations (\ref{eq:ME_EOM_ground})--(\ref{eq:tensor_prod_def}), we use a transition operator which is transformed consistently with the Hamiltonian.
To achieve this, we express the unitary transformation as the exponential of an anti-Hermitian generator: $U=e^{\Omega}$, with $\Omega^{\dagger}=-\Omega$~\cite{Morris2015}.
Any operator $\mathcal{O}^{\lambda}$ can then be consistently transformed by
\begin{equation}
\begin{aligned}
\bar{\mathcal{O}}^{\lambda} &= e^{\Omega}\mathcal{O}^{\lambda}e^{-\Omega} \\
&= \mathcal{O}^{\lambda} + [\Omega,\mathcal{O}^{\lambda}] + \frac{1}{2}[\Omega,[\Omega,\mathcal{O}^{\lambda}]] + \ldots
\end{aligned}
\label{eq:TensorMagnus}
\end{equation}
where we again use the scalar-tensor commutators of (\ref{eq:define_C}).

In the formulas presented in Appendix~\ref{sec:APP_products}, transition operators are assumed to be normal-ordered with respect to the reference $|\Phi_0\rangle$.
If $\mathcal{O}^{\lambda}$ is initially a one-body operator with $\lambda\neq0$, then this requires no additional work.
If $\mathcal{O}^{\lambda}$ has a two-body component---as is the case if we include meson-exchange currents, or if the bare operator has been SRG evolved in free space---then we need the formula for obtaining the normal-ordered form (indicated $\mathcal{N}^{\lambda}$) of $\mathcal{O}^{\lambda}$:
\begin{equation}
\begin{aligned}
\mathcal{N}_{ij}^{\lambda} = \mathcal{O}_{ij}^{\lambda}
+ \sum_{aJJ'}n_a(-1)&{}^{j_a+j_i-J'-\lambda} 
\sj{J}{J'}{\lambda}{j_j}{j_i}{j_a}\mathcal{O}_{iaja}^{JJ'\lambda},
\\
\mathcal{N}^{JJ'\lambda}_{ijkl} &{}= \mathcal{O}^{JJ'\lambda}_{ijkl}.
\label{eq:tensor_NO}
\end{aligned}
\end{equation}
This may be obtained by beginning with the usual $m$-scheme formula~\cite{Hergert2016} and applying (\ref{eq:CGID1}).
Here $n_a$ is the occupation fraction of orbit $a$, defined so that $0\leq n_a \leq 1$.

\subsection{\label{subsec:VS_Formalism}Valence space IMSRG}
In the valence-space (VS) formulation of the IMSRG, the unitary transformation $U$ decouples a valence space Hamiltonian $H_{\textrm{VS}}$ from the remainder of the Hilbert space (the excluded space) $H_{\textrm{excl}}$,
\begin{equation}
\bar{H}=UHU^{\dagger} = \bar{H}_{\textrm{VS}} + \bar{H}_{\textrm{excl}}.
\end{equation}
The eigenstates are obtained by a subsequent diagonalization of $\bar{H}_{\textrm{VS}}$ within the valence space.

The expectation value of $\mathcal{O}^{\lambda}$ between initial state $| \psi_i \rangle$ and final state $| \psi_f \rangle$ may be obtained by combining the matrix elements of $\mathcal{O}^{\lambda}$ with the one- and two-body transition densities (working with the consistently-transformed valence-space operators and wave functions)
\begin{equation}
\begin{aligned}
\langle \psi_f \| \mathcal{O}^{\lambda}\| \psi_i \rangle
&\rule{0pt}{8pt}= O_{0\textrm{b}} \delta_{fi} + \sum_{pq} O^{\lambda}_{pq}\text{OBTD}_{pq}^{\lambda}(\psi_f,\psi_i)\\
&\hspace{-1em}+ \frac{1}{4}\sum_{pqrs}\sum_{J_1J_2}\unnorm{O}^{J_1 J_2 \lambda}_{pqrs}\text{TBTD}_{pqrs}^{J_1 J_2 \lambda}(\psi_f,\psi_i).
\end{aligned}
\label{eq:VSexpectation}
\end{equation}
The one-body transition densities are defined by
\begin{equation}
\text{OBTD}_{pq}^{\lambda}(\psi_f,\psi_i) \equiv  \frac{\langle \psi_f \| [a^{\dagger}_p\times \tilde{a}_q]^{\lambda} \| \psi_i\rangle}{\sqrt{2\lambda+1}}
\end{equation}
and the two-body transition densities are
\begin{equation}
\text{TBTD}_{pqrs}^{J_1 J_2 \lambda}(\psi_f,\psi_i) \equiv  \frac{\langle \psi_f \| [A^{\dagger J_1}_{pq}\times \tilde{A}^{J_2}_{rs}]^{\lambda} \| \psi_i\rangle}{\sqrt{2\lambda+1}}.
\end{equation}
There is a clear parallel between (\ref{eq:VSexpectation}) and (\ref{eq:ME_EOM_ground}), due to the fact that the amplitudes $X_{ai}^J(\nu)$ and $\unnorm{X}_{abij}^{(J_1J_2)J}(\nu)$ correspond to the one- and two-body transition densities, respectively, between $|\Psi_\nu\rangle$ and the ground state.
For all the valence space results presented here, the diagonalizations were performed with the shell model code NuShellX@MSU~\cite{Brown2014}.
As NuShellX does not provide functionality to calculate the two-body transition densities for spherical tensor operators, an additional code has been developed~\cite{nutbar}.

For open-shell nuclei, we use the ensemble normal ordering (ENO) approach presented in Ref.~\cite{Stroberg2017}.
After the valence space is decoupled, we change the normal ordering reference to be the core of the valence space, which requires the use of (\ref{eq:tensor_NO}).

We note that the only approximation made in this procedure is the truncation to normal-ordered two-body operators.
Of course, the quality of this approximation depends on the choice of reference and valence space.

\section{\label{sec:Results}Results}
For all of the calculations presented here, with the exception of the results in section~\ref{sec:E3}, we employ the chiral NN interaction of Entem and Machleidt~\cite{Entem2003} at N$^{3}$LO with a cutoff $\Lambda_{NN}=500$ MeV, and the local 3N interaction of Refs.~\cite{Navratil2007,Gazit2009,Roth:2012qf} at N$^{2}$LO with a cutoff $\Lambda_{3N}=400$ MeV. We use an additional three-body energy truncation $E_{3max} \equiv e_1 + e_2 +e_3 \leq 14$, where $e_i = 2n_i + l_i$ corresponds to the $i$th single particle shell in the harmonic oscillator basis.  
The interactions are consistently SRG evolved~\cite{Bogner2007,Roth2014b} to a scale $\lambda_{SRG}=2.0~\fmi$.
This interaction has been shown to give an excellent reproduction of the binding energies in the vicinity of the oxygen isotopes \cite{Hergert:2013ij,Cipollone2013,Cipollone2015}, but it produces radii which are too small by roughly 10\% \cite{Lapoux2016a}.
Since we consider $E2$ transitions and moments, and the $E2$ operator goes as $r^2 Y^{(2)}$, we might expect quadrupole moments and $B(E2)$ strengths to be too small by 20\% and 35\%, respectively.
However, because these observables are dominated by the particles near the Fermi surface, while the radii are a bulk property, it is not obvious that this naive scaling should actually apply.

In most of the figures presented in the following, we present an observable calculated for various values of model space truncation $e_{max}$ and basis frequency $\hbar\omega$.
If the result is converged with respect to the model space truncation, it should not change as $e_{max}$ is increased, and it should be independent of $\hbar\omega$, corresponding to a horizontal line in our figures.

\subsection{Center-of-mass factorization\label{subsec:COM}}
Before presenting results for electromagnetic moments and transitions, we investigate the role of center-of-mass motion for our calculations.
The structure of self-bound nuclei is governed by a translationally-invariant Hamiltonian, which is why we expect factorization of the intrinsic and center-of-mass (c.m.) components of the wave function:
\begin{equation}\label{eq:com_fact}
|\Psi\rangle =  |\psi\rangle_{\textrm{in}} \otimes |\psi\rangle_{\textrm{c.m.}}.
\end{equation}
This is particularly important for our current investigation because we do not use translationally-invariant transition operators $\mathcal{O}^{\lambda}$ in order to avoid the inclusion of cumbersome recoil corrections~\cite{Eisenberg1970}.
If the c.m. wave function has angular momentum $\Lambda_{\textrm{c.m.}}=0$, then by the Wigner-Eckart theorem,
\begin{equation}
\langle \psi_{\textrm{c.m.}}(\Lambda_{\textrm{c.m.}}=0) | \mathcal{O}^{\lambda}_{\textrm{c.m.}} |  \psi_{\textrm{c.m.}}(\Lambda_{\textrm{c.m.}}=0) \rangle = 0,
\end{equation}
and there is no error incurred by including the c.m. part of the operator.
The IMSRG is formulated in a lab-frame harmonic oscillator basis with a truncation on the single particle energies ($2n+l\leq e_{max}$), and consequently we cannot ensure rigorous factorization of the c.m. and intrinsic wave functions.
We seek instead to demonstrate approximate factorization and, if necessary, project out spurious c.m. contamination. 

\subsubsection{Calculation of $H_{\textrm{c.m.}}$}
The form of the c.m. Hamiltonian is taken to be that of a harmonic trap, with the zero-point energy removed:
\begin{equation}\label{eq:Hcm}
H_{\textrm{c.m.}} (\tilde{\omega}) =  \frac{{\bf P}^2}{2 m A} +  \frac{1}{2} m A \tilde{\omega}^2 {\bf R}^2 - \frac{3}{2}\hbar \tilde{\omega}.
\end{equation}
We can compute properties of the c.m. wave function in a manner similar to the discussion in Refs.~\cite{Hergert2016,Hagen2009a,Hagen:2010uq}.
If the center-of-mass wave function is a Gaussian with oscillator length $b$, then it will have
\begin{equation}\label{eq:RcmandPcm}
\begin{matrix}
\langle R_{\textrm{c.m.}}^2 \rangle = \frac{3}{2} b^2, &\textrm{and} &
\langle P_{\textrm{c.m.}}^2 \rangle = \frac{3}{2} \frac{\hbar^2}{b^2}
\end{matrix},
\end{equation}
which implies
\begin{equation} \label{eq:RcmtimesPcm}
\xi_{\textrm{c.m.}} \equiv \sqrt{\langle R_{\textrm{c.m.}}^2 \rangle 
\langle P_{\textrm{c.m.}}^2 \rangle }/\hbar - \frac{3}{2} = 0.
\end{equation}
The deviation of $\xi_{\textrm{c.m.}}$ in (\ref{eq:RcmtimesPcm}) from zero indicates the deviation of the c.m. wave function from a pure Gaussian.
Once the Gaussian form is confirmed, the appropriate trapping frequency $\hbar\tilde{\omega}$ may be obtained from (\ref{eq:RcmandPcm}), with ${b^2 = \hbar/Am\tilde{\omega}}$ or, equivalently,
\begin{equation}
\hbar\tilde{\omega} = \frac{4}{3}\langle T_{\textrm{c.m.}}\rangle.
\end{equation}
\begin{figure}
\includegraphics[width=\columnwidth]{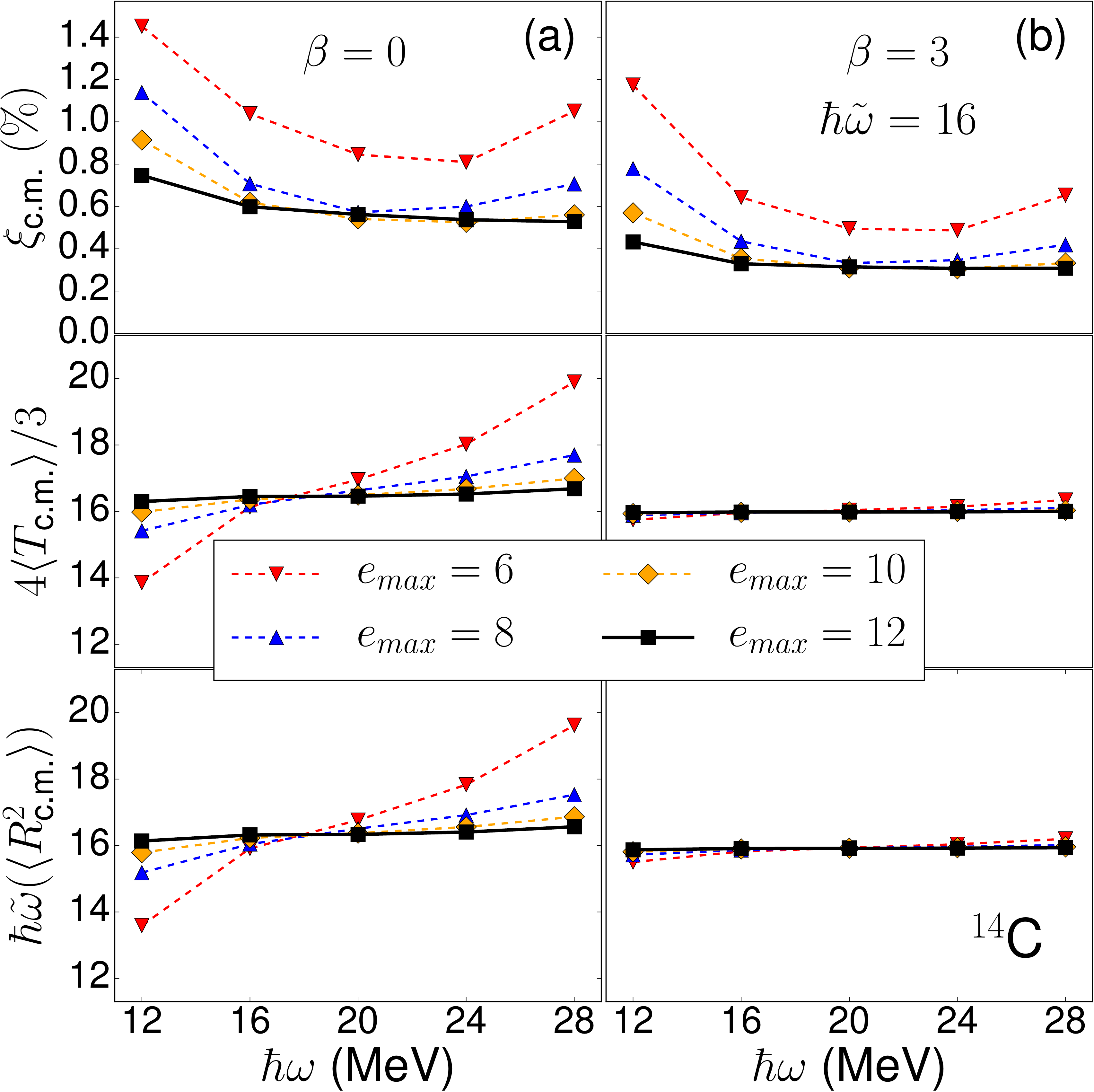}
\caption{The quantity $\xi_{\textrm{c.m.}}$, which gives an indication of the degree to which the c.m. wave function is a Gaussian (see text), calculated for the ground state of $^{14}$C. Also plotted are two methods of estimating the trapping frequency $\hbar\tilde{\omega}$ (MeV). Column (a) is without a c.m. trap, while column (b) is with $\beta=3$ at a frequency $\hbar\tilde{\omega}=16$ MeV.
}
\label{fig:C14_Com_RP}
\end{figure}
Figure~\ref{fig:C14_Com_RP} shows $\xi_{\textrm{c.m.}}$ results from IMSRG ground-state calculations for $^{14}$C.
Also shown are two ways of estimating $\hbar\tilde{\omega}$ from the expectation values of $T_{\textrm{c.m.}}$ and $R_{\textrm{c.m.}}^2$.
The right column of Figure~\ref{fig:C14_Com_RP} shows the same quantities, but with a c.m. trap (as described in the next section) with $\beta=3$ and $\hbar\tilde{\omega}=16$ MeV.
Clearly, the trap makes the c.m. wave function more Gaussian, though not perfectly Gaussian, and it speeds up the convergence of the c.m. wave function. 

\subsubsection{Treatment for excited states}
Spurious excited states manifest as nearly degenerate intrinsic states in nuclear spectra.   
\begin{figure}
\includegraphics[width=\columnwidth]{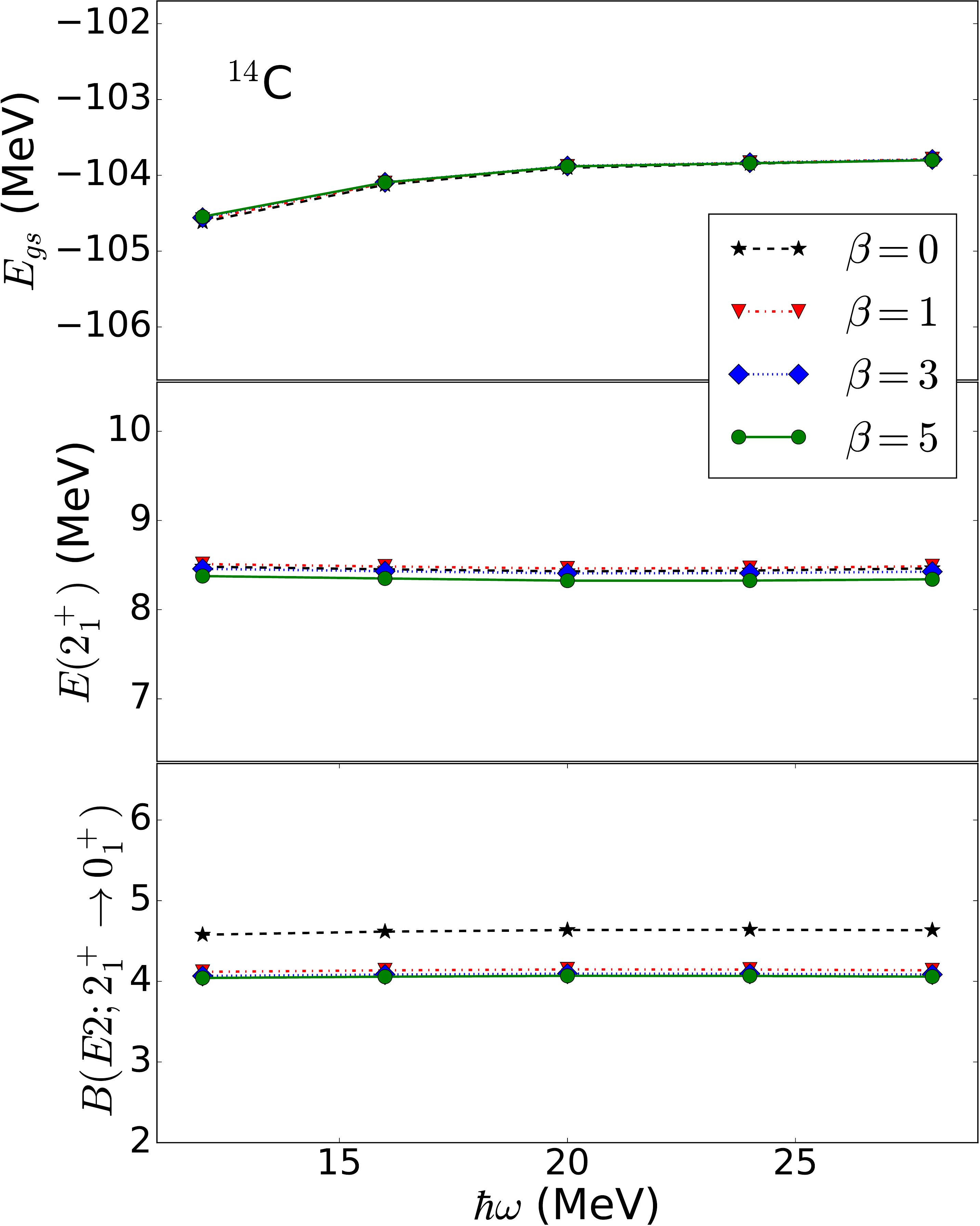}
\caption{Ground state, $2^+_1$ excitation energy, and $B(E2)$ values (in e$^2$fm$^4$) calculated at several values of the Lawson-Gloeckner scaling parameter $\beta$, for $^{14}$C with the EOM-IMSRG.}
\label{fig:C14_lawson}
\end{figure}
These states can be removed via the Lawson-Gloeckner method \cite{Gloeckner1974}, where the  intrinsic Hamiltonian is augmented with a scaled center-of-mass trap of the form of eq.~\ref{eq:Hcm},
\begin{equation}\label{eq:Law_Ham}
H =  H_{\textrm{in}} +\beta H_{\textrm{c.m.}}.
\end{equation}
Here, the scale factor $\beta$ can be taken to arbitrarily large values if sufficient factorization is achieved in calculations using $H_{\textrm{in}}$ only.  This process effectively shifts spurious states out of the spectrum by adding a large c.m. excitation energy. 

Figure~\ref{fig:C14_lawson} demonstrates this procedure  for $^{14}$C, for the ground state, first 2$^+$ excited state, and  $B(E2)$ value.  Quantities are calculated with the EOM-IMSRG(2,2) method. The energies are approximately independent of $\beta$, which may be taken naively as evidence of factorization for these states.  However, the $B(E2)$ value undergoes a sudden downward shift as the Lawson-Gloeckner term is introduced, but it saturates eventually and displays $\beta$-independence as we go to  higher $\beta$.  
Of course, the quadrupole operator is more sensitive to structural details of the wave function than the energy, and since we do not use it in a translationally-invariant form, it is not surprising that the $B(E2)$ value would be affected by the imperfect factorization of the wave functions. The fact that we eventually obtain a $\beta$-independent result suggests that the Lawson-Gloeckner method is an adequate alternative to explicitly including recoil corrections in the operator~\cite{Eisenberg1970}.

 Table~\ref{tab:C14_Ecm} gives the computed $E_{\textrm{c.m.}}$ for calculations with and without explicit inclusion of a center-of-mass trap via the Lawson-Gloeckner term.    
\begin{table}[h]
\centering
\caption{$E_{\textrm{c.m.}}$ for intrinsic ground state and first 2$^+$ state of $^{14}$C, computed at $e_{max}$=14 and $\hbar \omega$ = 20 MeV with EOM-IMSRG(2,2). Values are given for calculations using $H_{\textrm{in}}$ ($\beta$=0), and $H_{\textrm{in}}+\beta H_{\textrm{c.m.}}$ ($\beta$=1).}
\label{tab:C14_Ecm}
\begin{tabular*}{0.48\textwidth}{@{\extracolsep{\fill}}ccc}
\hline \hline
  $\beta$ & $E_{\textrm{c.m.}}(0^+_{gs})$ (MeV)  & $E_{\textrm{c.m.}}(2^+_{1})$ (MeV) \\
  \hline
  0  & 0.099 & 1.298 \\
  1 &  0.068 & 0.046 \\ \hline\hline
 \end{tabular*}
\end{table}
We expect a perfectly factorized wave-function to have $E_{\textrm{c.m.}}$=0 MeV, since our choice of $H_{\textrm{c.m.}}$ ensures that the c.m. ground state has zero energy.  For either case, the ground state wave function demonstrates limited contamination from spurious c.m. excitations, with $E_{\textrm{c.m.}} <$ 100 keV. The 2$^+$ state of $H_{in}$ does not exhibit this level of factorization, with $E_{\textrm{c.m.}}$=1.298 MeV, indicating a small admixture of spurious states. This level of contamination is ostensibly negligible for excitation energies, but evidently has important effects when the state is probed by the quadrupole operator. 
When the c.m. trap is explicitly added, $E_{\textrm{c.m.}}(2^+_1)$ is diminished to below 100 keV and accordingly, we see a shift in the $B(E2)$ value which corresponds to a recoil correction.
For the results presented below, we have checked and found that $^{14}$C is the only system where the c.m. trap has a noticeable effect.

\subsection{The deuteron\label{subsec:deuteron}}
As a first illustration, we consider ground-state properties of the deuteron.
This is useful for a few reasons. First, the system consists of only two particles and so induced three-body forces are irrelevant.
Further, the reference is taken to be the true vacuum, so the neglected three body forces do not feed back into the two body terms.
We should therefore expect the IMSRG(2) to be exact.
Second, full configuration interaction (FCI) calculations are easily performed for modest model spaces, allowing a direct evaluation of the precision of the IMSRG transformation.
Finally, we may treat the deuteron in the $0s$ valence space where the bare quadrupole moment is identically zero.
In this case, any non-zero quadrupole moment we obtain is entirely due to effects of the IMSRG evolution.
\begin{figure}
\includegraphics[width=\columnwidth]{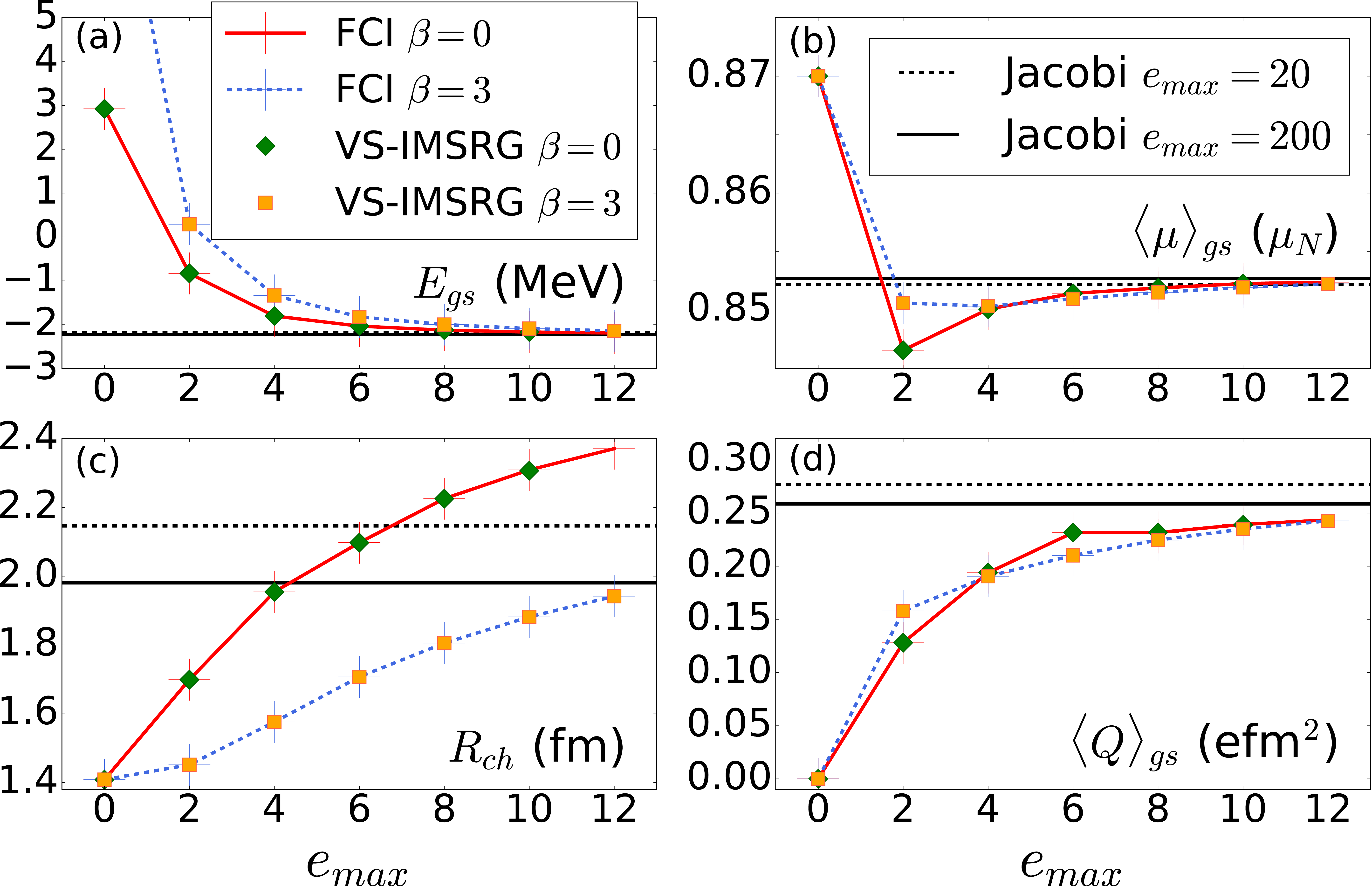}
\caption{Ground state properties of the deuteron calculated with a full diagonalization (labeled FCI), compared to the same properties calculated in the $0s$ valence space using operators transformed with the IMSRG. Also shown is the result obtained by diagonalizing in a Jacobi basis with $e_{max}=20$ and 200, in order to gauge convergence. Note that VS-IMSRG and FCI values are nearly the same.}
\label{fig:deuteron}
\end{figure}
Figure \ref{fig:deuteron} shows the ground-state energy, root-mean-square charge radius, quadrupole moment and magnetic moment of the deuteron, computed both with FCI and using the IMSRG to decouple the $0s$ valence space, followed by a trivial diagonalization.
We can see that the IMSRG calculation indeed reproduces the FCI.

Here again we see the effect of c.m. spuriosities in the deuteron wave function. While the energy and dipole moment converge to the exact values with little alteration from c.m. contamination, the charge radius overshoots it drastically. Although we have not reached convergence for the charge radius, it is evident that Lawson-Gloeckner scaling significantly reduces its value.
To get a sense of the rate of convergence for these observables in an oscillator basis, we have performed calculations in a relative Jacobi basis, where it is possible to go much higher in $e_{max}$.
We observe that the charge radius converges slowly in the Jacobi basis as well. 

\subsection{$p$-shell nuclei: comparison with NCSM\label{subsec:C14}}
The deuteron is, of course, an exceptionally simple case, due to the fact that there is not really a ``medium'', and so the IMSRG is really a free-space SRG evolution.
Once additional particles are considered, the NO2B approximation is used, and the IMSRG is no longer exact.
To test this approximation, we consider $p$-shell nuclei which may also be treated in the no-core shell model (NCSM). For these calculations, we use the same input Hamiltonian and include the 3N force completely, without using the NO2B approximation\footnote{Errors from the NO2B approximation in NCSM calculations will be different from those in IMSRG(2) calculations, as additional NO2B errors accumulate during the IMSRG(2) flow due to induced many-body forces.}
The NCSM calculations are presented as a function of the truncation parameter $N_{max}$ which limits the total number of oscillator quanta allowed above the minimum value.
For the $A=14$ systems, the $N_{max}=8$ results have been obtained using an importance truncation~\cite{Roth2007}.
We note that in the NCSM, the c.m. factorization is exact for any $N_{max}$ truncation.

\begin{figure}
\includegraphics[width=\columnwidth]{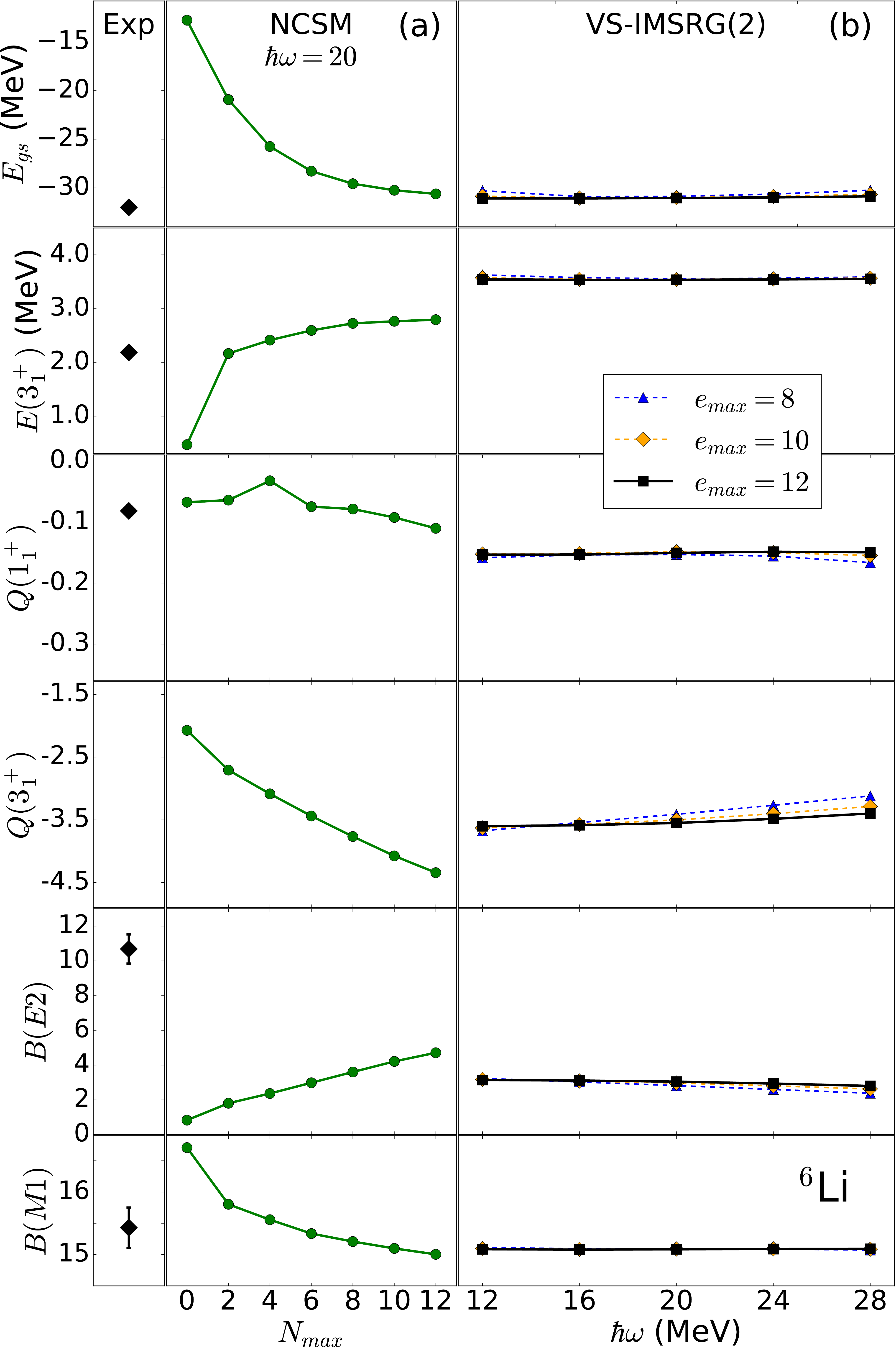}
\caption{Convergence of the ground-state energy, first $3^+$ excitation energy, quadrupole moments (in $e\fm^2$), $B(E2;3^+_1\rightarrow 1^+_1)$ ($e^2\fm^4$), and $B(M1;0^+_1\rightarrow 1^+_1)$ ($\mu_n^2$) of $^{6}$Li. The VS-IMSRG method (column (b))  is compared with NCSM results (column (a)) and experiment \cite{AJZENBERGSELOVE1984,Stone:2005moments,Pritychenko2016}. 
}
\label{fig:Li6_convergence}
\end{figure}

We begin by considering $^{6}$Li, which was previously studied in Ref.~\cite{Navratil1997} in the context of consistently-transformed electromagnetic transition operators using the Okubo-Lee-Suzuki method.
Figure \ref{fig:Li6_convergence} presents several observables for $^{6}$Li, calculated with the valence-space IMSRG, compared to NCSM and experiment.
We first observe that there is overall good agreement between the VS-IMSRG and NCSM, as well as with experiment, for the energy and quadrupole moment of the ground state.
In Ref. \cite{Lisetskiy2009}, where an effective $p$-shell $E2$ operator was obtained via an Okubo-Lee-Suzuki transformation, the small ground-state quadrupole moment was found to be the result of cancellations between the one and two-body pieces of the effective $E2$ operator.
We find a similar effect in this work\footnote{Since the IMSRG and Okubo-Lee-Suzuki transformations are not identical, there is no requirement that the breakdown into one- and two-body operators be the same in both approaches.}, though even greater in magnitude -- for example, for the $e_{max}=12$, $\hbar\omega$=20 calculation we find $Q_{1\textrm{b}}=-0.454$~$e$b and $Q_{2\textrm{b}}=0.301$~$e$b.
The results for observables involving the unbound $3^+$ excited state converge much more slowly in the NCSM, indicating missing continuum effects.
Such effects could be included using the NCSM with continuum~\cite{Hupin2015,Romero-Redondo2016}, but for our present concerns, this is unnecessary. Despite the importance of continuum effects, the VS-IMSRG(2) converges rapidly for observables involving the $3^+$ state. This indicates that errors incurred through the NO2B truncation hide the effects of the continuum. This produces excellent convergence properties by mistake; the VS-IMSRG(2) converges to an incorrect result without continuum degrees of freedom.  

A striking disagreement is found between experiment and the calculations of the $B(E2;3^+_1\rightarrow 1^+_1)$ strength.
As we will see, this will be a recurring observation.
Finally, we note that the $M1$ transition strength displays reasonable convergence and good agreement with experiment.

Another interesting case in the $p$ shell is $^{6}$He, which in the naive shell model consists of two neutrons outside a $^{4}$He core.
In this picture, any electric multipole observables are identically zero because all valence particles are electrically neutral.
This problem has historically been addressed by the introduction of an effective charge for the neutrons~\cite{Bohr1969}.
As in the deuteron case, $^{6}$He therefore allows us to test how the IMSRG evolution incorporates physics from outside of the valence space into the evolved operator, building up an effective charge in the process.

\begin{figure}
\includegraphics[width=\columnwidth]{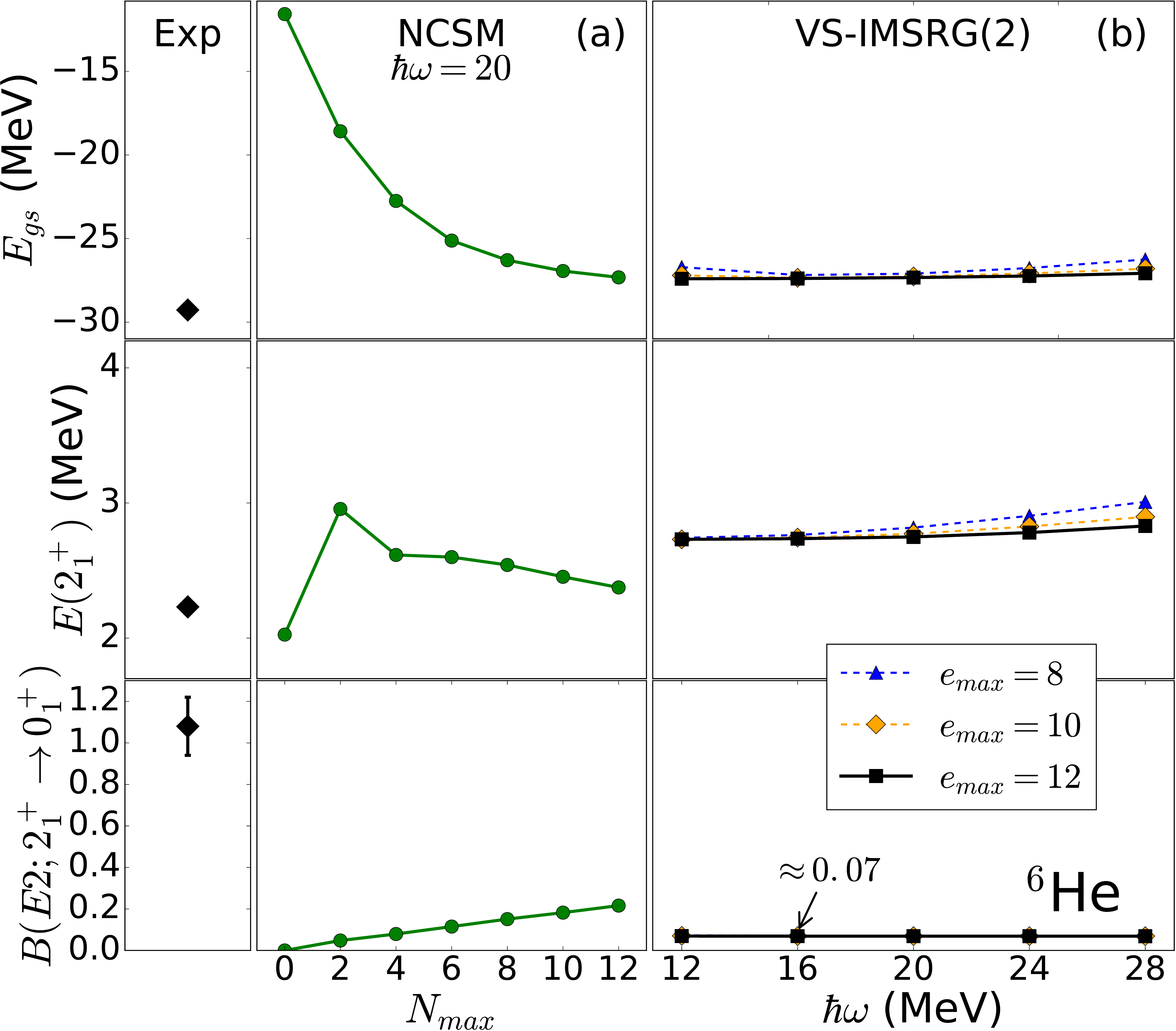}
\caption{Convergence of the ground-state energy, first $2^+$ excitation energy, and $B(E2)$ (in $e^2\fm^4$) to the ground state of $^{6}$He. Again, VS-IMSRG (column (b)) is compared with NCSM (column (a)) and experiment \cite{Pritychenko2016}. }
\label{fig:He6_convergence}
\end{figure}

Figure \ref{fig:He6_convergence} shows the results of VS-IMSRG and NCSM calculations for the ground-state energy, $2^+$ excitation energy, and $B(E2;2^+_1\rightarrow 0^+_1$) for $^{6}$He.
Like $^{6}$Li, the excited states of this nucleus are unbound, and in addition,
the $^{6}$He ground state can be characterized as a two-neutron halo~\cite{Zhukov1993}, which is difficult to describe in a truncated oscillator basis.
Nevertheless, we see that the ground state energy displays excellent agreement between the VS-IMSRG, NCSM, and experiment.
There is reasonable agreement as well for the energy of the $2^+$ state, although the NCSM result is not converged with respect to $N_{max}$ (again likely reflecting missing continuum effects).
However, for the B(E2), there is serious disagreement between all three.
The NCSM result is much lower than the experimental value, and shows no sign of convergence with respect to $N_{max}$.
This is perhaps not surprising, as the $E2$ operator is of long range, and therefore more sensitive to the halo effects.
The VS-IMSRG result appears converged with respect to $e_{max}$, but is smaller than the NCSM result as well as experiment---the latter by a factor of approximately 15---indicating that the NO2B approximation is insufficient in this case.

\begin{figure}
\includegraphics[width=\columnwidth]{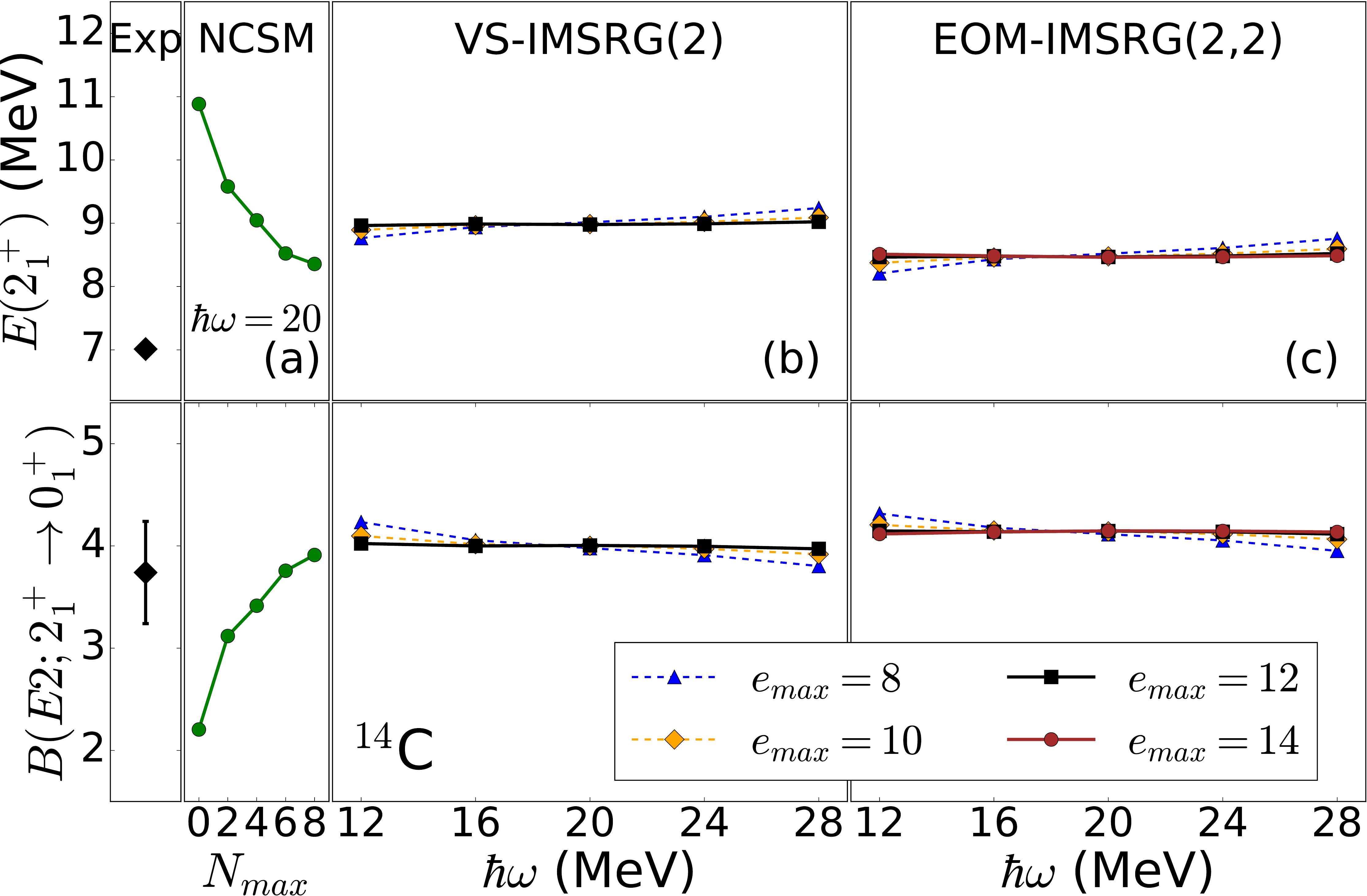}
\caption{Convergence of the first $2^+$ excitation energy and $B(E2)$ (in $e^2\fm^4$) to ground state of $^{14}$C. VS- and EOM-IMSRG methods (columns (b) and (c) respectively) are compared with NCSM (column (a)) and experiment \cite{Pritychenko2016}. }
\label{fig:C14_E2_convergence}
\end{figure}
As a third test in the $p$ shell, we consider $^{14}$C.
Because this is a closed-shell nucleus, we may employ the EOM-IMSRG as well as the VS-IMSRG, and a system of 14 particles is still feasible with the NCSM.
Figure \ref{fig:C14_E2_convergence} displays results for the $2^+_1$ excitation energy and $B(E2;2^+_1\rightarrow 0^+_1)$ for $^{14}$C.
Here, we find excellent agreement between NCSM and both variants of the IMSRG. We remind the reader that the IMSRG calculations are performed with an explicit center-of-mass trap, as in eq.~\ref{eq:Law_Ham}, using $\beta=1.0$ for $^{14}$C. This treatment only serves to remove spurious c.m. contamination of the $2^+_1$ state.

Of note are the excellent convergence properties of the IMSRG calculations. For the EOM-IMSRG, observables are nearly independent of the specified $\hbar \omega$ for the single-particle basis. VS-IMSRG calculations have not used the exhaustive model spaces of the EOM-IMSRG, but they too demonstrate desirable convergence features. The NCSM has begun to show convergence at $N_{max}$=8, but extrapolation methods must be used to reveal fully converged values. Hence the utility of the IMSRG: For light nuclei such as $^{14}$C, convergence is obtainable without extrapolation, and for heavier nuclei, we expect to be able to identify convergence trends clearly enough to make extrapolation procedures relatively painless compared to the prohibitively large uncertainties one would incur when exact methods such as NCSM are used.
Of course, the effect of the additional NO2B approximation must be fully investigated.

\begin{figure}
  \centering
\includegraphics[width=\columnwidth]{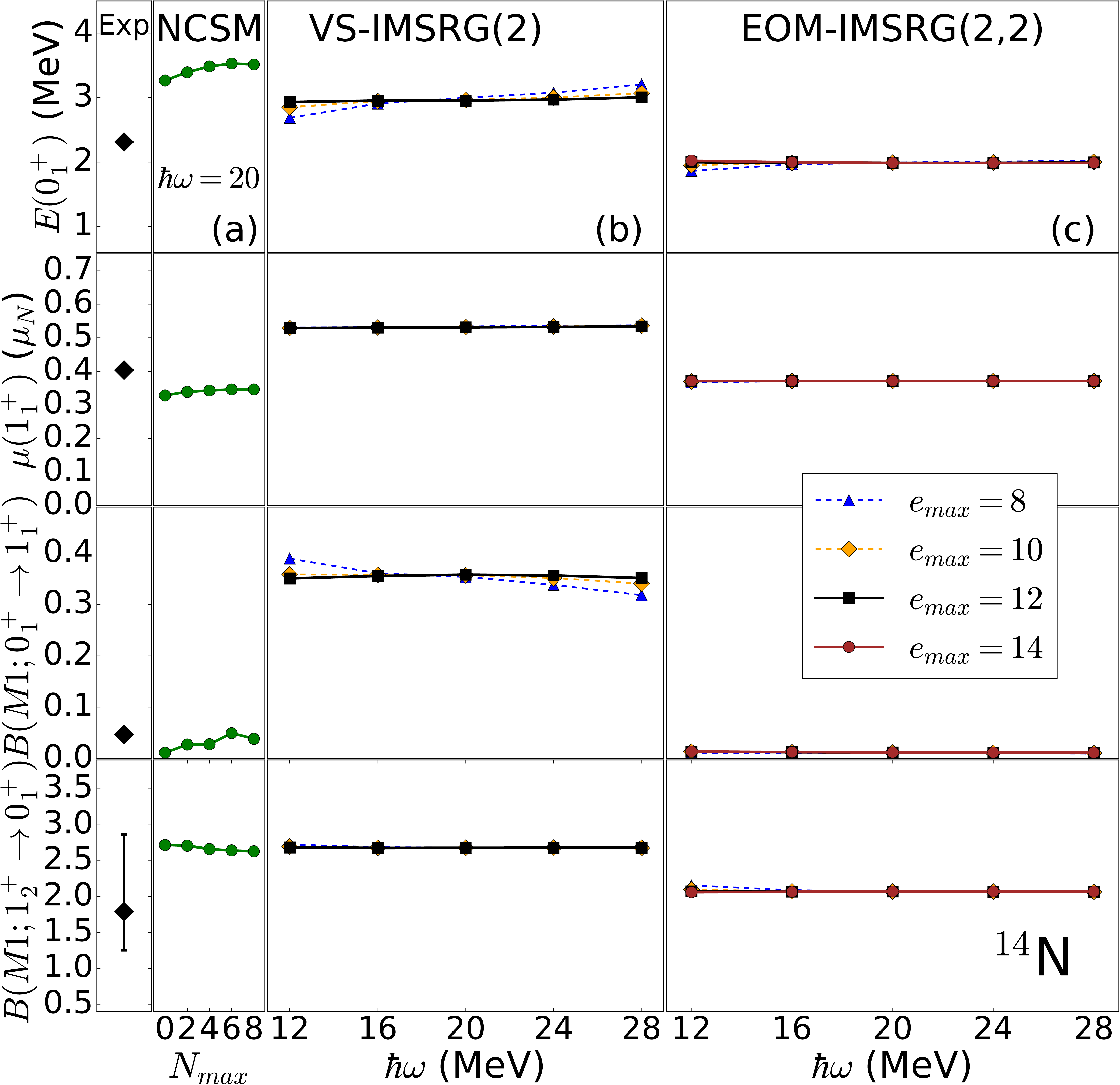}
\caption{Convergence of $0^+_1$ excitation energy, $B(M1)$ (in $\mu_N^2$) to ground state, and magnetic dipole moment of $^{14}$N. VS- and EOM-IMSRG methods (columns (b) and (c) respectively) are compared with NCSM (column (a)) and experiment \cite{AjzenbergSelove1991,Stone:2005moments}. }
\label{fig:N14_M1_convergence}
\end{figure}

As a final test in the $p$ shell, we analyze the isobaric neighbor nucleus $^{14}$N. Here the EOM-IMSRG requires the use of a charge-exchange formalism, i.e., ladder operators which exchange one neutron for a proton. Figure~\ref{fig:N14_M1_convergence} displays the $0^+_1$  excitation energy for $^{14}$N, the ground state magnetic dipole moment, and the $M1$ transition strengths $B(M1;0^+_1\rightarrow 1^+_1)$ and $B(M1;1^+_2\rightarrow 0^+_1)$.
The agreement among methods is moderate, with the exception of the transition $B(M1;0^+_1\rightarrow 1^+_1)$ to the ground state.
We note that this relatively weak transition, which is an analog of the Gamow-Teller beta decay of $^{14}$C, was found to result from a subtle cancellation between various contributions~\cite{Maris2011,Ekstrom2014a}, so that small errors on an absolute scale appear large on a relative scale.
Regardless, the disagreement between VS-IMSRG and EOM-IMSRG will be investigated in the future.

\subsection{$sd$ and $fp$ shell systems}
Ultimately, the power of IMSRG approaches to excited states and effective operators will be the ability to describe these properties in medium- to heavy-mass regions where exact methods are not computationally tractable. In this section we investigate the quality of these calculations for several medium-mass nuclei, again using the electric quadrupole and magnetic dipole operators as case studies. 

\subsubsection{Electric quadrupole observables}
\begin{figure}[ht]
  \centering
  \includegraphics[width=\columnwidth]{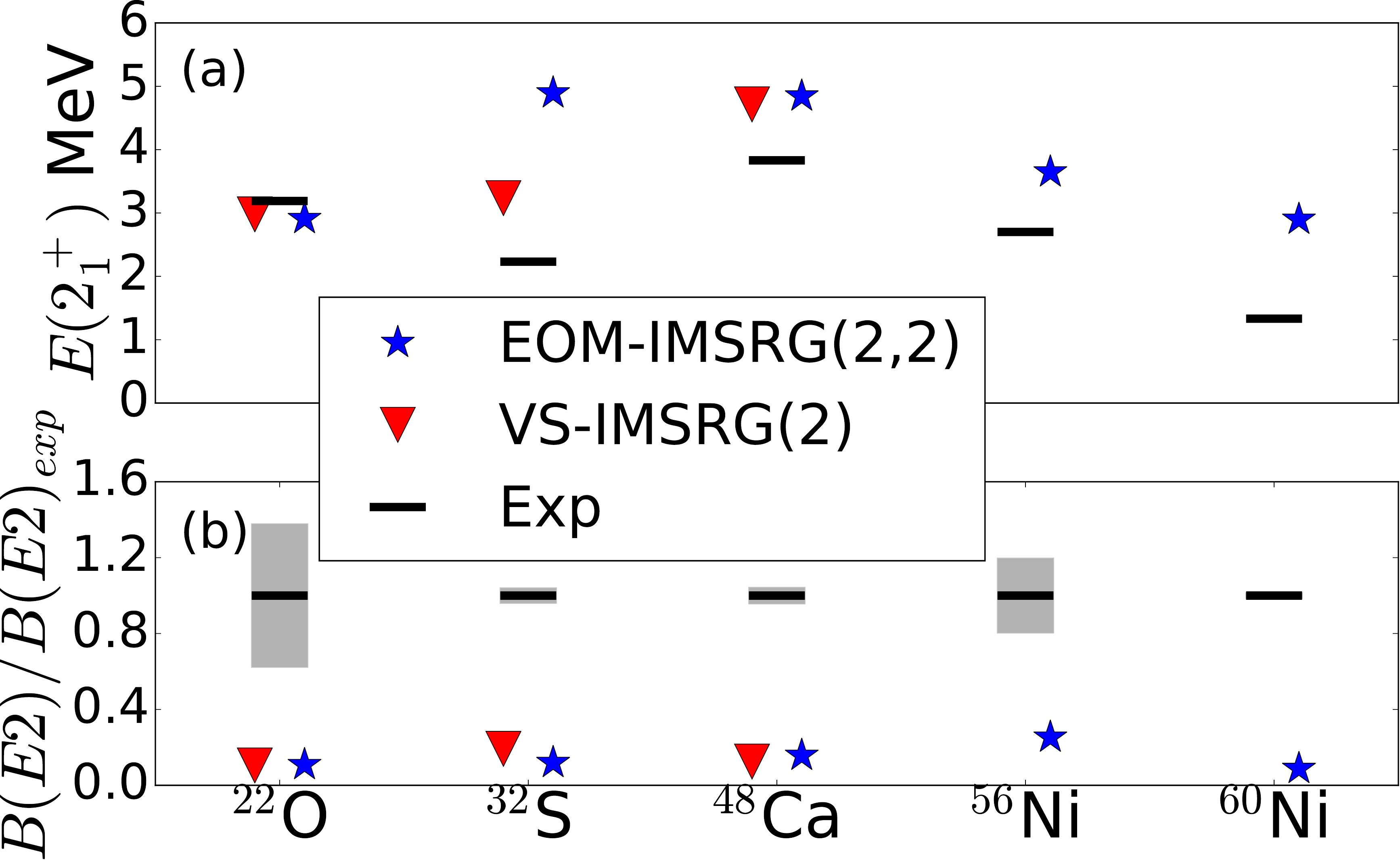}
  \caption{Results of EOM-IMSRG(2,2) and VS-IMSRG(2) calculations of the $2^+_1$ excitation energy (a), and the $B(E2;2^+_1\rightarrow 0^+_1)$ value (b) for several closed-shell nuclei in the $sd$ and $pf$ shells. Due to experimental values that vary by several orders of magnitude, the $B(E2)$ values are scaled such that experiment is unity. Computations are performed at $\hbar \omega=20$ MeV and $e_{max}=12$. Experimental results are taken from~\cite{Pritychenko2016}. }
  \label{fig:E2_summary}
\end{figure}

Figure~\ref{fig:E2_summary} displays the first 2$^+$ excitation energies and $B(E2;2^+_1\rightarrow 0^+_1)$ strengths for several nuclei in the $sd$ and $pf$ shells.  We find excellent convergence properties, as we did in the $p$ shell, and we see reasonable agreement with experiment for the excitation energies. However, transition strengths are generally underpredicted by an order of magnitude. These results are strikingly consistent between the two methods. A tentative explanation for the diminished strength in $^{22}$O and $^{48}$Ca is provided by the lack of valence protons. In order to describe the transition in these nuclei, valence neutrons must be dressed consistently as quasi-neutrons possessing an effective charge.

The absence of any appreciable strength in the two IMSRG calculations appears to be convincing evidence that IMSRG evolutions, when restricted to the two-body operator level (i.e., VS-IMSRG(2) and EOM-IMSRG(2,2)), do not sufficiently renormalize the neutron charges.
However, this discrepancy is evident in many nuclei, regardless of shell structure; we see the same underpredictions in $^{32}$S, and $^{56,60}$Ni, which lie in middle of their respective major shells, with plenty of valence protons to model an electromagnetic transition. 

Table~\ref{tab:E2_compiled} compiles the results from several of the calculations presented here, where $B(E2)$ corresponds to $B(E2;2^+_1 \rightarrow 0^+_{1})$.
In the far right column, we include the the Weisskopf estimate for the transition~\cite{Blatt1979}.
The Weisskopf estimate, given by
\begin{equation}\label{eq:weisskopf}
B(E2)_W = \frac{9 r_0^4}{100 \pi} A^{4/3} e^2 \textrm{fm}^4, 
\end{equation}
models the transition as a single proton excitation from a core with the empirical nuclear radius $r_0 A^{1/3}$, where $r_0 = 1.2$ fm. Excitations that are dominated by a single 1p1h transition will yield experimental $B(E2)$ values near the Weisskopf estimate. This picture certainly falls short of describing those nuclei with magic proton numbers, such as $^{22}$O, but it is nonetheless instructive to consider what the single particle estimates are for even these nuclei, as they describe neutrons with an effective charge in this case.  

\begin{table}[h]
\centering
\caption{$E2$ transition strengths from first excited $2^+$ state to $0^+$ ground state for even-even nuclei (in $e^2fm^4$). Experiment \cite{Pritychenko2016} and Weisskopf~\cite{Blatt1979} single particle estimates are compared with IMSRG calculations.}
\label{tab:E2_compiled}
\begin{tabular*}{0.49\textwidth}{@{\extracolsep{\fill}}cD{,}{}{-1}D{.}{.}{-1}D{.}{.}{-1}D{.}{.}{-1}}
\hline \hline
\rule{0pt}{8pt}
  Nucleus &
  \multicolumn{1}{c}{$B(E2)_{exp}$}  &   \multicolumn{1}{c}{$B(E2)_{EOM}$} & 
\multicolumn{1}{c}{$B(E2)_{VS}$} &
\multicolumn{1}{c}{$B(E2)_{W}$}\\
  \hline
  \rule{0pt}{10pt}
  $^6$He  & 1,.1(1)  &  & 0.07 & 0.6   \\
  $^{14}$C  & 3,.6(6)& 4.1 & 3.9 & 2.0 \\
    $^{22}$O  & 4,.2(1.6)& 0.5 & 0.4 & 3.7 \\
  $^{32}$S  & 59,(1) & 7.2  & 11.3 & 6.0\\
    $^{48}$Ca  & 17,(2) & 2.6 & 2.0 & 10.4\\	
$^{56}$Ni  & 91,(17) &  30.7 &  & 12.7\\
  $^{60}$Ni  & 186,(3) & 16.2 &  & 14.0\\
  \hline\hline
 \end{tabular*}
\end{table}

We find that computed $B(E2)$ values track with Weisskopf estimates rather than actual experimental values, except in the case of a magic proton shell closure, where computations are significantly smaller than the Weisskopf estimates, suggesting that indeed, the renormalization of neutron effective charges may not be sufficient in our IMSRG calculations.
Moreover, the fact that many of the experimental $B(E2)$ values are significantly larger than the single particle estimates indicates that collectivity which is neglected by VS-IMSRG(2) and EOM-IMSRG(2,2) calculations may be more critical to $E2$ transition strengths than it is to excitation energies. 

\begin{table}[hb]
\caption{Effective charges for the $E2$ operator, obtained by decoupling the $sd$ shell with a reference of $^{17}$O for neutrons ($\delta e_{\nu}$) and $^{17}$F for protons ($\delta e_{\pi}$), and taking the ratio with the bare matrix elements for protons. }
\label{tab:eff_charge}
\begin{tabular*}{0.49\textwidth}{@{\extracolsep{\fill}}ccD{.}{.}{-1}D{.}{.}{-1}}
\hline
\hline
$a$ & $b$ & \multicolumn{1}{c}{$\delta e_{\nu}$} &\multicolumn{1}{c}{ $\delta e_{\pi}$} \\
\hline
$0d_{5/2}$ & $0d_{5/2}$ & 0.213 & 0.026 \\
$0d_{5/2}$ & $0d_{3/2}$ & 0.248 & 0.075 \\
$0d_{5/2}$ & $1s_{1/2}$ & 0.184 & 0.039 \\
$0d_{3/2}$ & $0d_{3/2}$ & 0.120 &-0.003\\
$0d_{3/2}$ & $1s_{1/2}$ & 0.111 &-0.007 \\
\hline
\hline
\end{tabular*}
\end{table}

As a further illustration, we present in Table~\ref{tab:eff_charge} the orbit-dependent effective charges for the one-body piece of $E2$ operator, obtained in a VS-IMSRG(2) calculation of $^{17}$O.
Here, we define the effective charges so that
\begin{equation}
\begin{matrix}
e_{\pi} = 1 + \delta e_{\pi}, & & e_{\nu} = \delta e_{\nu}.
\end{matrix}
\end{equation}
The values listed correspond to a model space truncation $e_{max}=12$, and a basis frequency of $\hbar\omega=20$ MeV.
(The bare matrix elements are evaluated in the Hartree-Fock basis, so these results are essentially independent of the basis frequency).
We obtain a neutron effective charge of approximately 0.1--0.2, considerably smaller than the standard phenomenological value of 0.5. 
We repeat the exercise for the proton effective charge, using $^{17}$F as the reference, and we obtain very small (and even negative) values of $\delta e_{\pi}$.
This discrepancy between proton and neutron effective charges is similar to the effect seen in second order perturbation theory in Ref.~\cite{Thoresen1998}, and will be investigated in a future work.

Another possible explanation for diminished $E2$ observables is deficiencies in the input interactions. As previously discussed, the NN+3N(400) interaction systematically underpredicts nuclear radii, which is tied to its inability to reproduce nuclear saturation. Since the electric quadrupole operator has the same radial dependence as the point-nucleon radius operator, we might naively expect an increase in predicted $B(E2)$ values when using an input interaction which properly reproduces radii, such as N$^{2}$LO$_{\text{sat}}$~\cite{Ekstrom2015}. We computed $B(E2)$s for the nuclei shown in fig.~\ref{fig:E2_summary} with this interaction, using EOM-IMSRG(2,2). We found that a small enhancement is indeed observed ($\sim$50\% increase), but N$^2$LO$_{\text{sat}}$ still systematically underpredicts $B(E2)$ values for these nuclei, indicating that while the interaction does play an important role, missing correlations are still likely to be a major source of error.

\subsubsection{Magnetic dipole observables}
We now turn to $M1$ observables, where the Weisskopf estimate (1.79 $\mu_N^2$) is independent of $A$, and we therefore expect the transition to have similar properties from nucleus to nucleus, unlike $E2$ observables. We have calculated $B(M1)$ values in $^{14}$C and $^{22}$O, where we have observed excellent consistency between VS- and EOM-IMSRG, as we did for $E2$ observables. Our predictions for $B(M1;1^+_1 \rightarrow 0^+_1)$ are in the vicinity of 1$\mu_N^2$ for both nuclei. The experimental value for $^{14}$C is 0.3938$\pm$0.0895$\mu_N^2$, a difference which could potentially be accounted for by missing meson-exchange currents in our dipole transition operators. 
\begin{figure}
  \centering
  \includegraphics[width=\columnwidth]{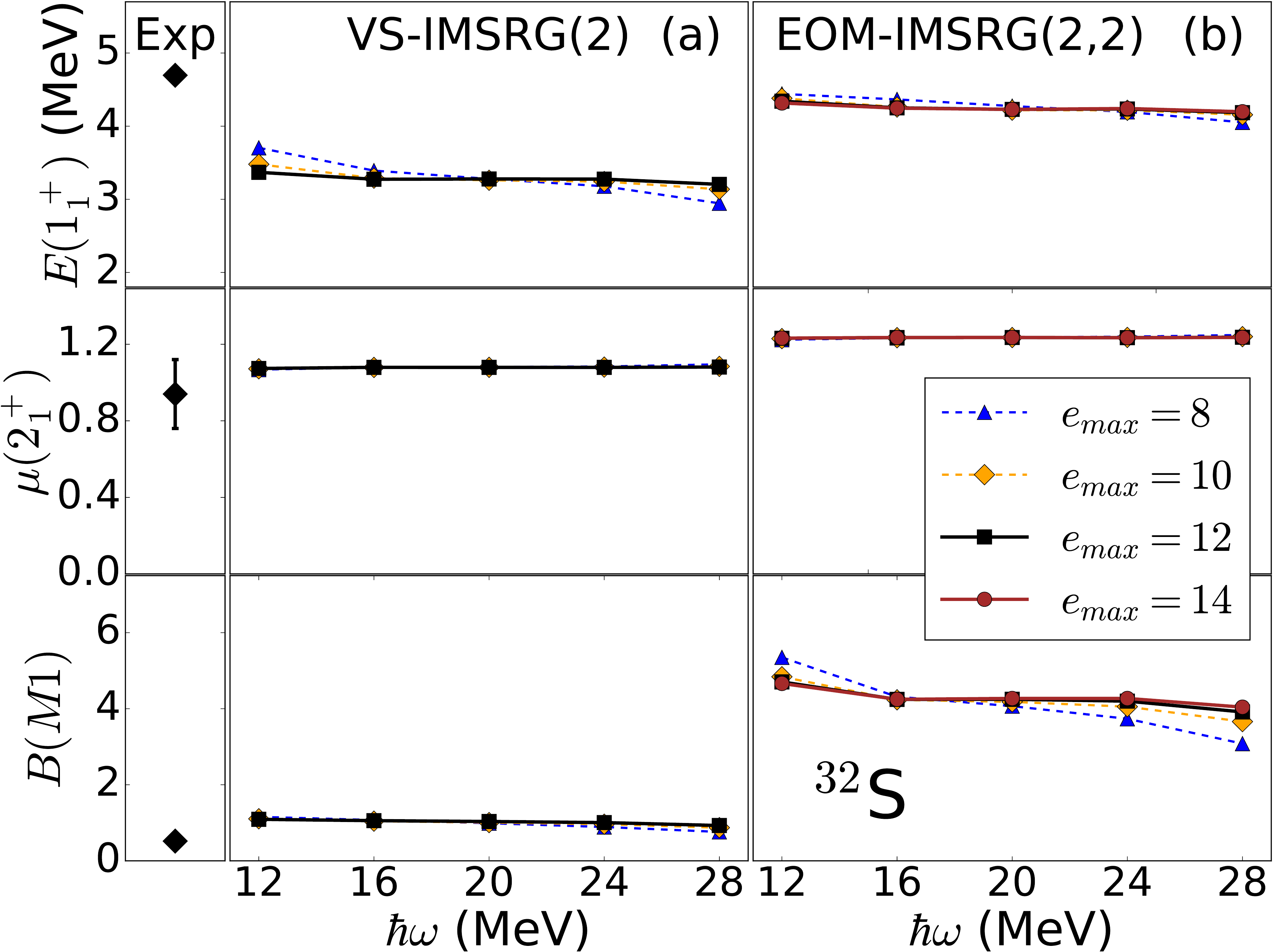}
  \caption{Convergence of the $1^+_1$ excitation energy, $B(M1;1^+_1 \rightarrow 0^+_1)$ (in $\mu_N^2$) and $2^+_1$ magnetic dipole moment ($\mu_N$) of $^{32}$S. VS-IMSRG (column (a)) and EOM-IMSRG (column (b)) methods are compared with experiment \cite{Stone:2005moments,KangasmakiBM1}.}
  \label{fig:S32_M1_convergence}
\end{figure}

We also compute $M1$ observables for the $1^+_1$ and $2^+_1$ states in $^{32}$S.
Figure~\ref{fig:S32_M1_convergence} shows results from these calculations.
We find good agreement between the methods for the magnetic moment of the $2^{+}$ state, and with experiment, which is on the order of the naive shell-model estimate.
There is some disagreement between the methods for the $B(M1)$ transition strength, which is three orders of magnitude smaller than the Weisskopf estimate.
As with the $M1$ transition in $^{14}$N, this is likely due to subtle cancellations, and the apparent error is amplified.
\begin{figure}
  \centering
  \includegraphics[width=\columnwidth]{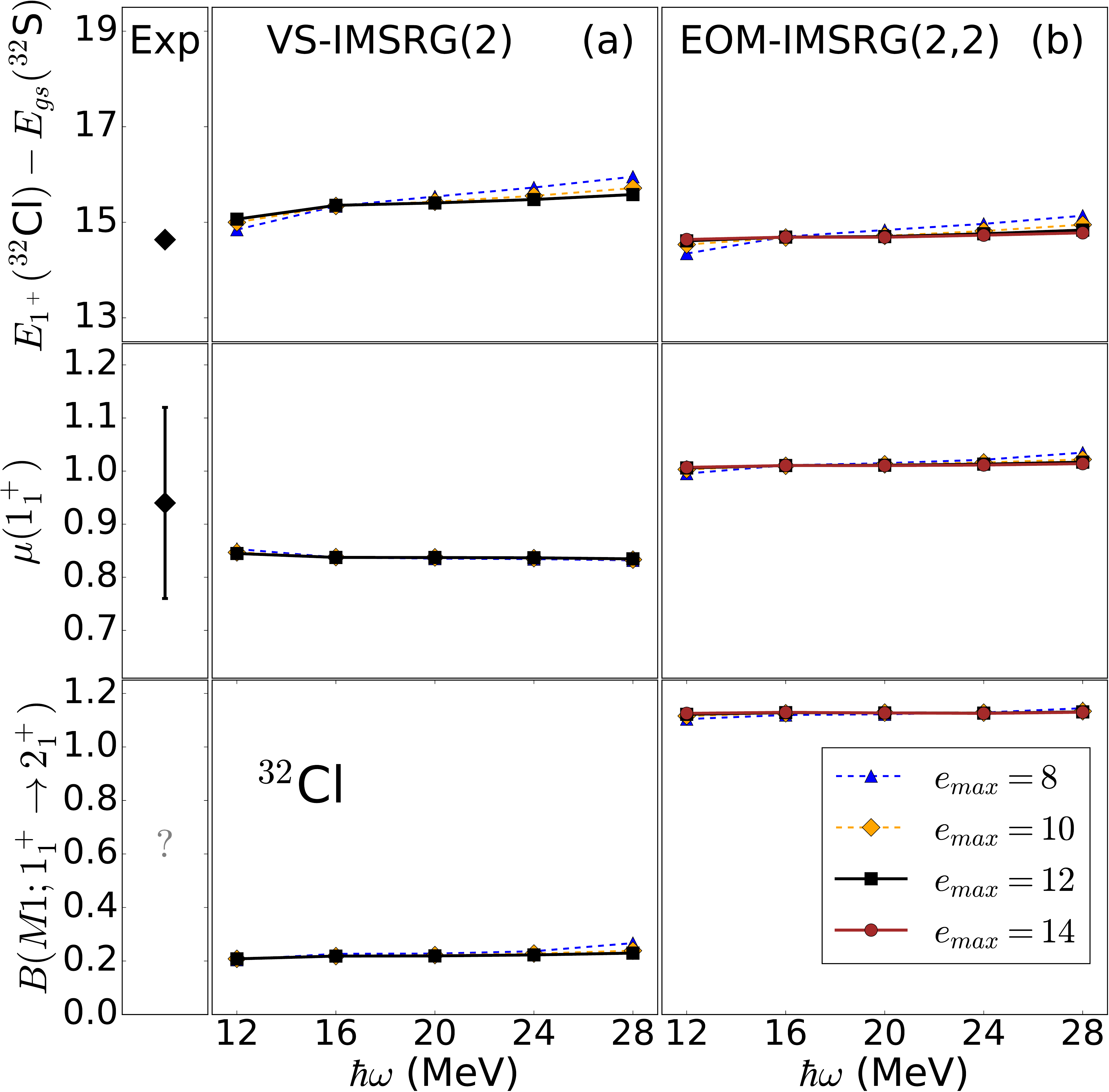}
  \caption{Energy and magnetic dipole moment (in $\mu_N$) of the $1^+_1$ state, and $B(M1)$ to the $2^+_1$ state ($\mu_N^2$) of $^{32}$Cl. VS-IMSRG (column (a)) and EOM-IMSRG (column (b)) methods are compared with experiment \cite{Stone:2005moments}, where available. 
  }
  \label{fig:Cl32_M1_convergence}
\end{figure}

In addition, we investigate $^{32}$Cl, whose ground state can be thought of as a charge-exchange excitation of $^{32}$S, where a neutron is exchanged for a proton. Experimentally, $^{32}$Cl is observed to have a $1^+$ ground state with a nearly degenerate $2^+$ state at $89.9$ keV~\cite{Gade:2004ow}. Both IMSRG methods fail to properly order these states with the employed interaction, instead producing a $2^+$ ground state with a $1^+$ excited state at 660 and 430 keV for VS- and EOM-IMSRG, respectively. 

Figure~\ref{fig:Cl32_M1_convergence} shows convergence of the energy and magnetic dipole moment of the $1^+$ state, as well as predictions for the $M1$ transition strength between the $1^+$ and $2^+$ states. The energy is given here as an excitation from the $^{32}$S ground state, as it is calculated in the EOM-IMSRG as an excited state of $^{32}$S with a charge-exchange excitation operator. Disagreement between EOM- and VS-IMSRG is more notable here than for other nuclei, though both methods show qualitative agreement with experiment where available. Particularly troubling is the disagreement in the $B(M1)$ value, which suggests a large discrepancy in the way higher-order correlations are incorporated into the $2^+$ state by the two methods.

To investigate this discrepancy, we attempt to approximately reconcile the different approximations made.
First, we restrict the VS-IMSRG calculation to allow only one proton and no neutrons in the $0d_{3/2}$ orbit, corresponding to the 1p1h part of the EOM-IMSRG ladder operator, and we obtain $B(M1)=1.35~\mu_N^2$.
Next, we allow two protons and one neutron in the $0d_{3/2}$ orbit, which incorporates all 2p2h EOM-configurations in the $sd$ shell, as well as some 3p3h configurations, and we obtain $B(M1)=0.41~\mu_N^2$.
Finally, we restrict the EOM-IMSRG calculation to only allow $sd$-shell configurations, and we obtain a minor suppression of $B(M1)=1.08~\mu_N^2$.
From this, we conclude that the structure of the $2^+$ state 
is sensitive to configuration mixing effects that are not sufficiently captured with 1p1h and 2p2h excitations out of $^{32}$S. 
 
We have computed the magnetic dipole properties of several nuclei, seeing reasonable consistency between EOM- and VS-IMSRG for most observables considered. Limiting ourselves to closed shell cases only, this corresponds to what is seen for $E2$ observables.
In order to compare more precisely with experiment, we should also include the effects of mesonic currents which occur within the nucleus during the transition. Work in that direction is underway.

\subsection{Electric Octupole Transitions\label{sec:E3}} 
\begin{figure}
\includegraphics[width=\columnwidth]{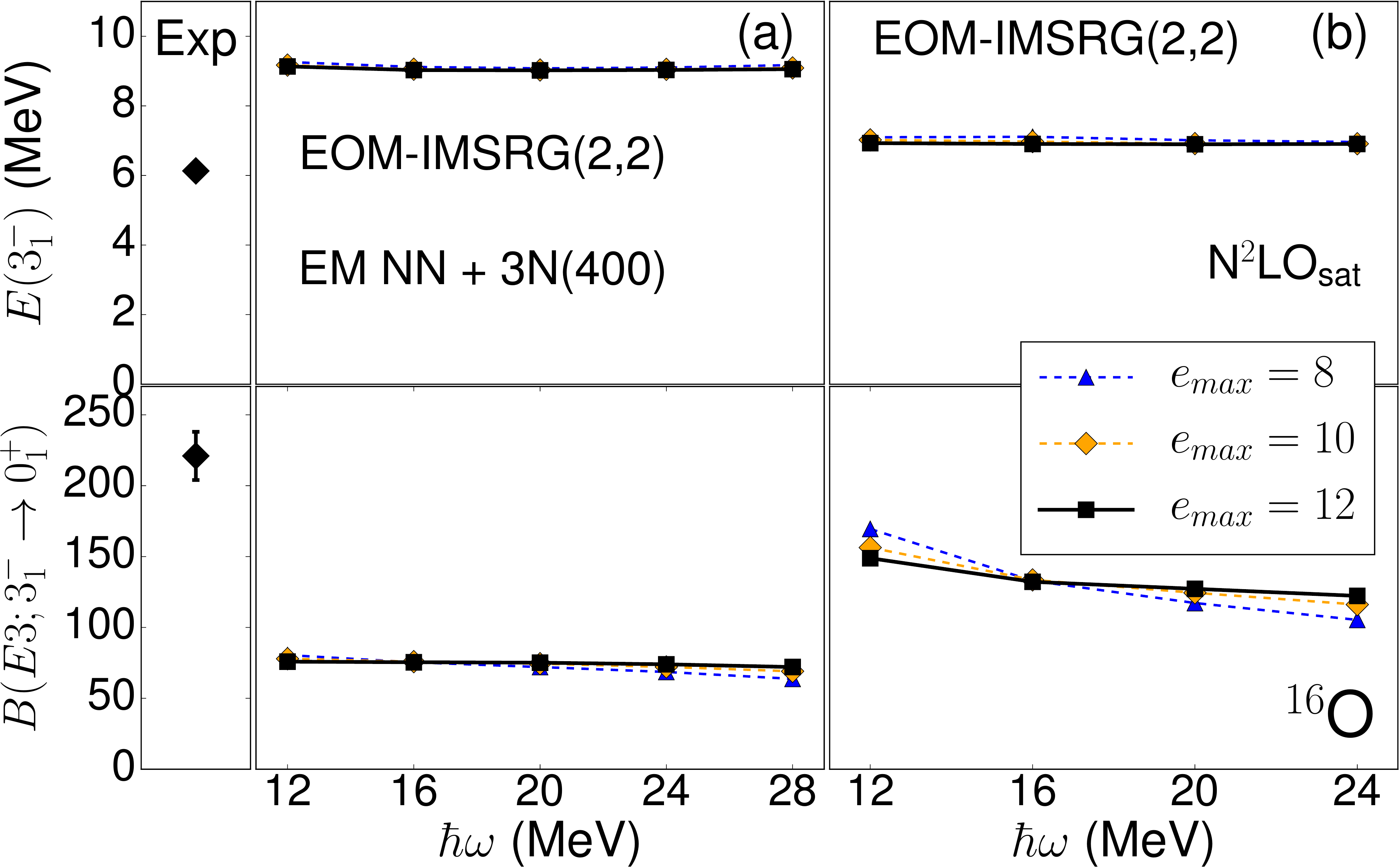}
\caption{Excitation energies of the first $3^-$ state of $^{16}$O, along with corresponding $B(E3)$ strength (in $e^2\fm^6$) for the transition to ground state. These values are computed using the EOM-IMSRG(2,2) with N$^2$LO$_{\textrm{sat}}$ (column (b)) and the Entem and Machleidt NN(500)-3N(400) interaction (column (a)), and are compared with experiment~\cite{Kibedi2002}.}
\label{fig:O16_E3_convergence}
\end{figure}

The electric octupole transition offers an additional test of the EOM-IMSRG.
(The VS-IMSRG is not currently able to decouple multi-shell valence spaces, and consequently cannot treat parity-changing operators).
We investigate the transition strengths from the first $3^-$ state to ground state for the doubly magic nuclei $^{16}$O and $^{40}$Ca. Figure~\ref{fig:O16_E3_convergence} shows the convergence of this calculation for $^{16}$O. This is an interesting case 
study, as the $3^-_1$ excitation energy has been shown to correlate with the $^{16}$O charge radius and thus depends on saturation properties of the interaction \cite{Ekstrom2015}. For this reason, we compare calculations with the NN+3N(400) interaction to those using N$^2$LO$_{\textrm{sat}}$, which is fit to the $^{16}$O charge radius \cite{Ekstrom2015}. We see an improvement of the excitation energy when using N$^2$LO$_{\textrm{sat}}$, moving from 9.03 MeV with the NN+3N(400) interaction to 6.90 MeV, in significantly better agreement with  the experimental value at 6.13 MeV. Both interactions underpredict the $B(E3)$ value for the transition to the ground state, with the saturating interaction showing greater strength than the NN+3N(400) interaction. Despite EOM partial norms indicating 90$\%$ 1p1h content in the $3^-_1$ wave-function, higher order correlations may play a significant role in the structure pertinent to the $E3$ transition, as $\alpha$-clustering may be important to the structure of the $3^-_1$ state~\cite{Kanada-Enyo:2016}. If this were true, EOM-IMSRG(2,2) would not be an appropriate approximation for such a state, and the suppressed $E3$ strength would be an expected result owing to the missing collectivity.

\begin{figure}
\includegraphics[width=\columnwidth]{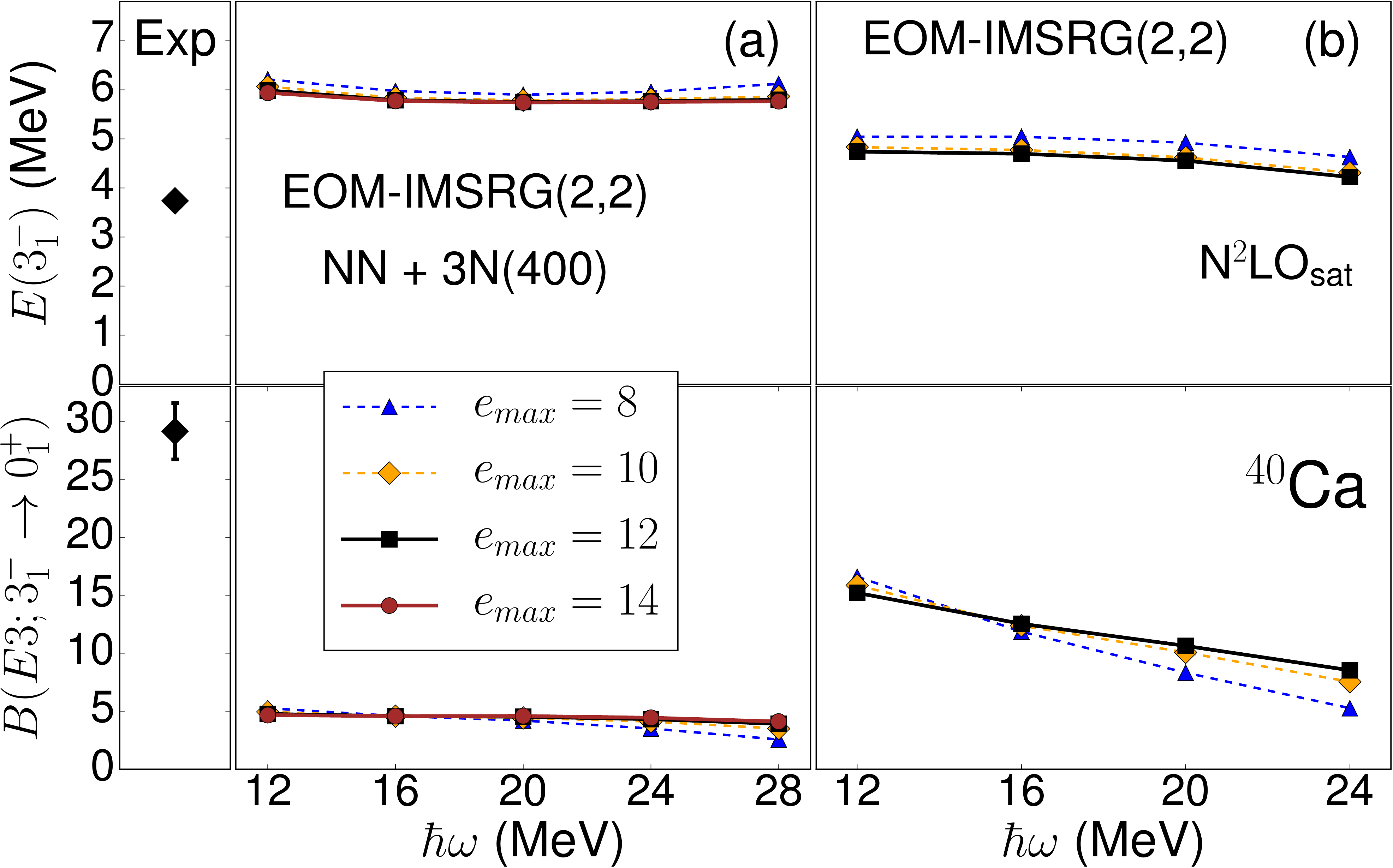}
\caption{Excitation energies of the first $3^-$ state of $^{40}$Ca, along with corresponding $B(E3)$ strength (in $10^2 e^2\fm^6$) for the transition to ground state. These values are computed using the EOM-IMSRG(2,2) with N$^2$LO$_{\textrm{sat}}$ (column (b)) and the Entem and Machleidt NN(500)-3N(400) interaction (column (a)), and are compared with experiment~\cite{Kibedi2002}.}
\label{fig:Ca40_E3_convergence}
\end{figure}

A similar picture presents itself for $^{40}$Ca in Figure~\ref{fig:Ca40_E3_convergence}, where again, N$^2$LO$_{\textrm{sat}}$ improves the excitation energy but underpredicts the $B(E3)$ strength. For either interaction, the discrepancy is less striking than that seen in $^{16}$O, but the deviation is significant nonetheless. Notable is the poor convergence features exhibited by N$^2$LO$_{\textrm{sat}}$, where results are seemingly dependent on the basis frequency $\hbar \omega$ for both energy and $B(E3)$ value. For  $pf$-shell nuclei and beyond, the $E_{3max}$ truncation has been shown to be a significant source of errors in many-body calculations~\cite{Hagen2016b}. This is certainly the case here (we note the $3^-$ state of $^{40}$Ca is composed of excitations into the $pf$ shell in a naive shell model picture), where varying $E_{3max}$ between 12 and 14 produces shifts in the excitation energy by 2-4 MeV, and shifts in the $B(E3)$ value by several hundred e$^2$fm$^6$ for N$^2$LO$_{\textrm{sat}}$. However, these errors are not as dramatic for the NN+3N(400) interaction, where the corresponding shifts are on the order of 10 keV and 10 e$^2$fm$^6$, for energies and $B(E3)$ values respectively, hence the more desirable convergence features.  

Computed $E3$ strengths suffer from largely the same shortcomings as $E2$ strengths, where we see a significant reduction of the strength from that of experiment. The Weisskopf single-particle estimates for $^{16}$O and $^{40}$Ca are 15.2 and 95.0 $e^2\fm^6$ respectively.
The immense size of the experimental values compared with these estimates indicates a strong level of collectivity in these $3^-_1$ states, which is apparently missing in our calculations, although computed $E3$ strengths are indeed larger than the single-particle estimates.  

\subsection{\label{sec:comp_cont}Comparing and contrasting methods}
While we have seen remarkable agreement between the VS-IMSRG(2) and EOM-IMSRG(2,2), there are some discrepancies in the predictions made by either method.
These discrepancies are the result of some combination of two sources of error:
The two methods decouple different sets of orbits---the EOM-IMSRG decouples a single reference determinant, while the VS-IMSRG decouples the valence space and core, i.e., multiple states at once---and this leads to different errors incurred by the NO2B approximation.
Typically, the VS-IMSRG requires a more substantial rotation and therefore is more susceptible to error, though in cases with a small gap above the Fermi surface the opposite may be true.
On the other hand, the EOM-IMSRG(2,2) lacks the ability to describe higher-order correlations in states with minimal 1p1h character.
This underscores the fact that the two methods are complementary, and different classes of states fall into the sets that are best described by either method. 

The VS-IMSRG takes into account all possible valence-particle configurations within the specified valence space. States that are described well by phenomenological shell-model approaches should then be described appropriately by the VS-IMSRG. As the shell model can describe collective properties such as deformation, states of this character are well described by this method, provided that their collectivity is restricted to the VS-IMSRG(2) decoupled valence space. 
On the other hand, states with significant contributions from multiple major shells, in particular unnatural parity states, are unreachable by this method in its current state. Methods to decouple a multi-shell valence space are still under investigation.

The EOM-IMSRG is not restricted by the core/valence space paradigm, but rather derives its computational simplicity from a restriction of the configurations included in the diagonalization.
Natural- and unnatural-parity states are therefore treated on the same footing;
however, any state will be poorly described if it is dominated by particle-hole excitations that are left out of the definition of the ladder operators.
For the EOM-IMSRG(2,2), we work with a space of 1p1h and 2p2h configurations.
In this case, states with strong 1p1h content with respect to the fully decoupled reference state are best described by the method. States with 2p2h dominant wave functions are accessible, but the ground-state-decoupled Hamiltonian still introduces strong correlations between these states and 3p3h excitations. For states with a relatively small admixture of 3p3h configurations, we expect that perturbative corrections will be sufficient for the inclusion of missing triples content \cite{Parzuchowski2017}. However, if the state couples strongly to triple excitations, a full EOM-IMSRG(3,2) treatment is required. Any state which has significant 4p4h or higher correlations in its transformed wave function would require EOM-IMSRG(4,2), and so on.

It is difficult to clearly determine a priori which method will perform best for any given state because, at present, there is no prescription to assign accurate theoretical error bars to these calculations.
However, one can make inferences about which method will perform best based on parity arguments, the number of valence nucleons, and the ``magnitude'' of the IMSRG transformation, indicated by the norm $\|\Omega\|$ (cf.~eq.~\eqref{eq:TensorMagnus}).
Here it will be important to develop a reliable measure of unitarity to quantify IMSRG(2) truncation errors associated with the aggressiveness of the decoupling scheme, e.g.,
targeting single or multiple states, enforcing additional block-diagonality for the Hamiltonian, etc.

\subsection{The effects of consistent operator evolution }
It is worth assessing the impact of consistently applying the IMSRG transformation to the operators discussed thus far; if the bare operators give essentially the same results, then this extra effort is unnecessary. 
By bare operators, we mean operators expressed in the Hartree-Fock basis, which have not been consistently evolved along with free-space SRG softening.
Because the interaction has been softened with the free-space SRG, the operator evolution is not exactly consistent in the first place.
Despite that caveat, free-space softening transformations are understood to have little effect on long-range operators such as the electromagnetic multipole operators discussed here, since the principal effect of SRG softening is to renormalize short-range physics.
Nonetheless, the problem is being given increasing attention in the nuclear physics community \cite{Anderson:2010br,Schuster:2014rg,More2015,Schuster2015}.

\begin{figure}[h]
\includegraphics[width=\columnwidth]{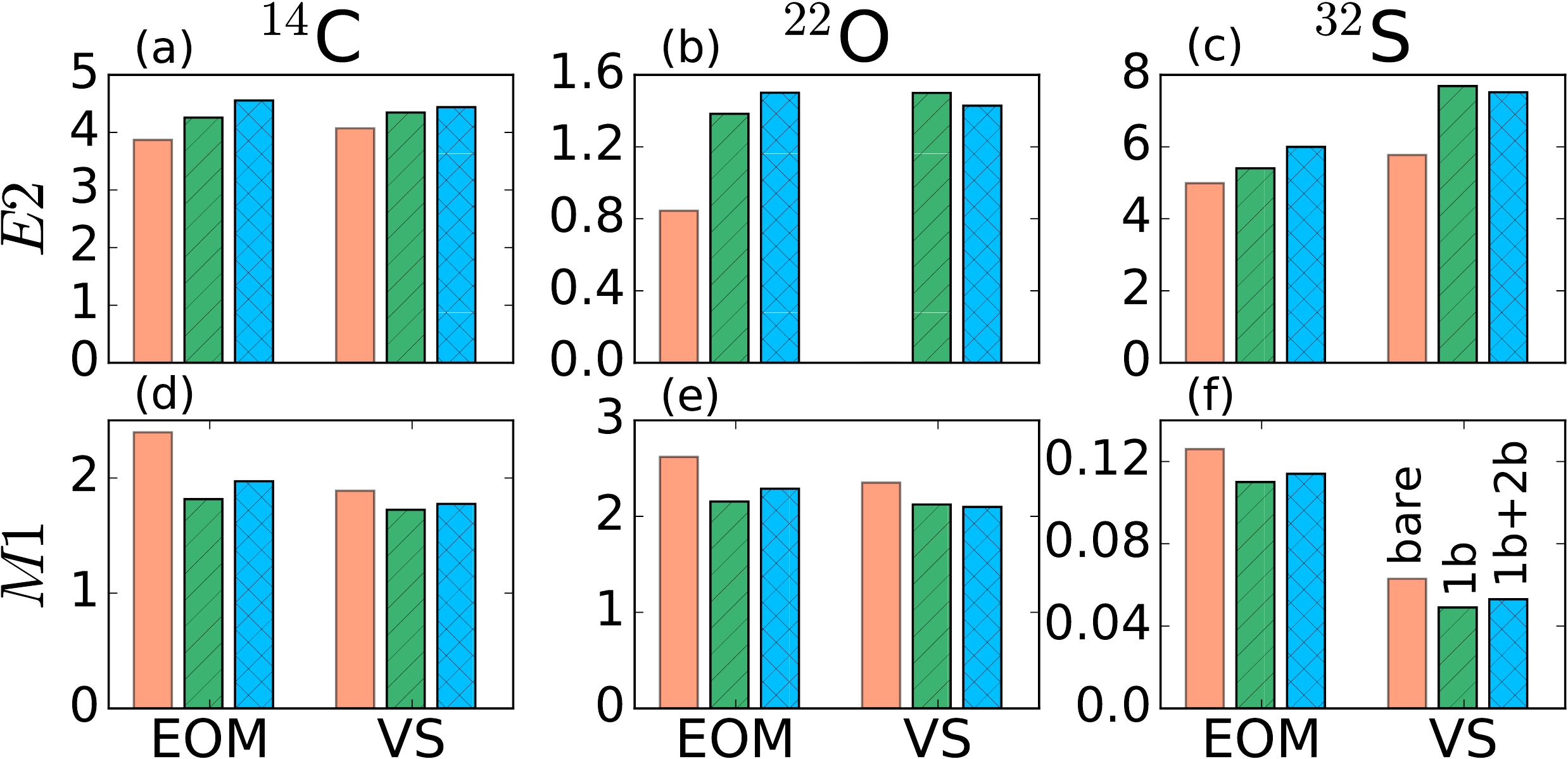}
\caption{Transition matrix elements $\langle 0^+_1 \| E2 \| 2^+_1\rangle$ and $\langle 0^+_1 \| M1 \| 1^+_1 \rangle$ computed for select nuclei using EOM-IMSRG(2,2). Calculations are performed for \emph{bare} operators (orange bars) and operators \emph{dressed} by consistent IMSRG evolution, both the one-body part (green single-hashed bars) and the one-body plus two-body part (blue double-hashed bars) . Values are expressed in $e~\fm^{2}$ and $\mu_N$ for $E2$ and $M1$ operators, respectively.}
\label{fig:operator_components}
\end{figure}

The IMSRG transformation is expected to have a noticeable effect on transition operators, as it renormalizes dynamic correlations in the nucleus, which are crucial to transition behavior.
Figure \ref{fig:operator_components} presents
a few examples of transition matrix elements computed with and without consistent evolution of the operator. 
$M^{1b}_{bare}$ refers to the reduced matrix element of the operator expressed in the Hartree-Fock basis without consistent evolution, (using wave functions computed with the evolved Hamiltonian), and $M^{1b,2b}_{dressed}$ refer to the same calculation with consistently evolved operators.
It is evident from these values that IMSRG evolution transforms a bare one-body operator into a many-body operator.

For EOM-IMSRG(2,2) calculations, the induced two-body term generally contributes less than 10\% of the total magnitude, suggesting that induced three-body terms (neglected in this work) should have an even smaller effect in many cases.
Consistently evolved $M1$ transition matrix elements exhibit a 10-20\% decrease in magnitude compared with the bare operator, and the equivalent comparison for $E2$ transitions show an increase in magnitude of $\sim$20\%, except in the case of $^{22}$O, where the magnitude increases by 77.5\%.

As the decoupling schemes of the VS- and EOM-IMSRG are different, we include results for both methods.
We note that the VS-IMSRG(2), despite employing a more substantial decoupling, produces a smaller two-body contribution to the matrix elements for all cases studied here.
In several cases, the two-body contributions in the VS-IMSRG destructively interfere with the one-body contributions, although the one-body part dominates.
The VS-IMSRG results demonstrate the critical effect of charge renormalization in $^{22}$O, which has no $sd$-shell protons, and thus vanishing strength when using the bare operator. 

From the results shown here, it is evident that consistent operator evolution is indeed important in \emph{ab initio} nuclear structure calculations.
The details of the operator evolution will of course depend on the system under consideration, and on the correlations a given solution method is able to describe.
For example, if a given transition is dominated by shell-model like configurations, then the effect of operator evolution should be small in the VS-IMSRG.
A similar argument applies for the EOM-IMSRG for transitions that are dominated by 1p1h contributions.
On the other hand, in some cases---such as $^{22}$O in the VS-IMSRG---the bare operator will give no contribution at all
and the effects of operator evolution are indispensable.

\section{\label{sec:Summary}Summary and Outlook}
In this work we have compared and contrasted two recently developed methods for the computation of excited-state properties and related observables.
The VS-IMSRG uses IMSRG decoupling to create an effective Hamiltonian for shell model diagonalization, while the EOM-IMSRG performs an approximate, particle-hole truncated diagonalization of the ground-state-decoupled Hamiltonian.
We applied both methods to computations of electromagnetic moments and transition strengths. 

It was observed that IMSRG ground-state decoupling approximately factorizes the center of mass (c.m.) component of excited state wave functions such that the effect of contamination on the energies is usually negligible.
This factorization is not always sufficient for excited states and electromagnetic strengths, which are sensitive to c.m. contamination due to the operators being expressed in lab coordinates.
We confirmed that the factorization can be improved by placing the system in a harmonic oscillator trap which only acts on the c.m. coordinate.

The EOM-IMSRG and VS-IMSRG give consistent results in the majority of cases analyzed.
While theoretical error bars are necessary for rigorous comparison, the methods qualitatively agree with each other and also with the NCSM, with a few noted exceptions.
This latter fact affords us confidence in the results of IMSRG excited-state calculations in heavier nuclei, for instance in the $sd$ and $pf$ shells. 

Notably, experimental $E2$ observables were underpredicted by roughly an order of magnitude in all nuclei except $^{14}$C.
Results of our calculations instead tracked with Weisskopf single-particle estimates, indicating that the inclusion of higher-order collective excitations will be critical for a proper description of $E2$ observables.
A thorough investigation of the treatment of these observables will be forthcoming.
$E3$ observables were computed for doubly magic nuclei, where a similar pattern was observed. 
$M1$ observables, while consistent between employed methods, showed differences from experiment that could potentially be accounted for by the inclusion of meson exchange currents~\cite{Pastore2013a} in future works.

In general, electromagnetic observables are well converged with respect to the size of the model space.
The main source of error in many of these calculations is evidently truncation errors associated with the NO2B approximation.
IMSRG applications to excited states will continue to improve as technical developments are made regarding truncation errors and decoupling strategies.
For example, the EOM-IMSRG can be improved significantly by the perturbative inclusion of 3p3h excitations \cite{Parzuchowski2017}, and the range of applicability of this method will be extended greatly upon extension to multireference formalism.
VS-IMSRG methods will continue to improve as strategies for decoupling cross-shell valence spaces are developed, enabling an explicit treatment of important degrees of freedom. 
Results should also improve as we devise strategies for the inclusion of neglected three-body operators that are induced by the IMSRG evolution.

\begin{acknowledgments}
We would like to thank Titus Morris, Jason Holt, Sonia Bacca, and Gaute Hagen for helpful discussions, and Angelo Calci for providing SRG evolved NN+3N matrix elements for the IMSRG calculations.
For performing the angular momentum coupling, we found Ref.~\cite{Wormer2006} to be especially helpful.
The valence-space IMSRG code used in this work~\cite{imsrg++} makes use of the Armadillo \texttt{C++} library \cite{Armadillo}.
TRIUMF receives
funding via a contribution through the National Research
Council Canada. 
This work was supported in part by
NSERC, the NUCLEI SciDAC Collaboration under the
U.S. Department of Energy Grant
DE-SC0008511, and the National Science Foundation under
Grants No.~PHY-1404159, PHY-1614130, and PHY-1614460.
Computing resources were provided through an allocation
at the J\"ulich Supercomputing Center, as well as  
Michigan State University's Institute for Cyber-Enabled Research (iCER).
PN received support from an INCITE Award on the Titan supercomputer at the Oak Ridge Leadership Computing Facility at Oak Ridge National Laboratory, and from Calcul Quebec and Compute Canada.
SRS would like to thank Kyle Leach for providing additional computational resources.
\end{acknowledgments}

\appendix
\section{Commutator relations in $m$-scheme}
In $m$-scheme, the commutation relations for tensor operators are identical to those for scalar operators.
For completeness, we give them here.
We seek to compute the commutator
$\opz = [\opx,\opy]$, where $\opx$, $\opy$, and $\opz$ have zero-, one- and two-body parts.
$\opz$ will also have a three-body part, which we neglect in the present discussion, in keeping with the NO2B approximation.
The components of $\opz$ are broken up into various contributions based on the particle rank of the terms in $\opx$ and $\opy$.
For example, (\ref{eq:Z110}) below indicates the contribution to the zero-body piece of $\opz$ by the one-body pieces of $\opx$ and $\opy$.
Additionally, to facilitate the later angular momentum coupling, (\ref{eq:Z222pphh}) and (\ref{eq:Z222ph}) are broken up into contributions involving particle-particle and hole-hole intermediates, as opposed to particle-hole intermediates.
\begin{equation}
\opz_0(11\rightarrow 0) = \sum_{pq}(n_p-n_q)\opx_{pq}\opy_{qp}
\label{eq:Z110}
\end{equation}
\begin{equation}
\opz_0(22\rightarrow 0) = \frac{1}{2}\sum_{pqrs}(n_pn_q\bar{n}_r\bar{n}_s)
\opx_{pqrs}\opy_{srpq}
\end{equation}
\begin{equation}
\opz_{pq}(11\rightarrow 1) = \sum_{r}(\opx_{pr}\opy_{rq}
-\opy_{pr}\opx_{rq})
\end{equation}
\begin{equation}
\opz_{pq}(12\rightarrow 1) = \sum_{rs}(n_r-n_s)
(\opx_{rs}\opy_{sprq} - \opy_{rs}\opx_{sprq} )
\end{equation}
\begin{align}
\opz_{pq}(22\rightarrow 1) = \sum_{rst}(n_rn_s\bar{n}_t &+ \bar{n}_r\bar{n}_sn_t) \notag \\
\times( \opx_{tprs}&\opy_{rstq} - \opy_{tprs}\opx_{rstq})
\end{align}
\begin{widetext}
\begin{equation}
\opz_{pqrs}(12\rightarrow 2) = \sum_{t}
(\opx_{pt}\opy_{tqrs} +
\opx_{qt}\opy_{ptrs} -
\opx_{tr}\opy_{pqts} -
\opx_{ps}\opy_{pqrt})
\end{equation}
\begin{equation}
\opz_{pqrs}(21\rightarrow 2) = -\sum_{t}
(\opy_{pt}\opx_{tqrs} +
\opy_{qt}\opx_{ptrs} -
\opy_{tr}\opx_{pqts} -
\opy_{ps}\opx_{pqrt})
\end{equation}
\begin{equation}
\opz_{pqrs}(22\rightarrow 2;pp/hh) = 
\frac{1}{2}\sum_{tu}(1-n_t-n_u)
(\opx_{pqtu} \opy_{turs} - \opy_{pqtu}\opx_{turs})
\label{eq:Z222pphh}
\end{equation}
\begin{equation}
\opz_{pqrs}(22\rightarrow 2;ph) =
-\sum_{tu}(n_t-n_u)(1-P_{pq})(1-P_{rs}) \opx_{puts}\opy_{tqru}
\label{eq:Z222ph}
\end{equation}

\section{Expressions for tensor-scalar commutator and tensor-tensor product}
\label{sec:APP_products}
The consistent evolution of effective spherical tensor operators, along with the computation of excited states using EOM-IMSRG formalism require expressions for the commutator of a scalar operator and a spherical tensor operator of arbitrary rank $\lambda$, given by
\begin{equation}
\mathcal{C}^{\lambda}_{\mu} \equiv 
 \mathcal{S}\mathcal{T}^{\lambda}_{\mu} - \mathcal{T}^{\lambda}_{\mu}\mathcal{S},
\end{equation}
where 
\begin{equation}
\begin{aligned}
C^{\lambda}_{pq} &= \sum_{a} \left( S_{pa} T^{\lambda}_{aq} - T^{\lambda}_{pa}S_{aq} \right) - \sum_{ab}(n_a-n_b)\Biggl(
  \bar{S}^{\lambda}_{p\bar{q}a\bar{b}} T^{\lambda}_{ab} -
  \hat{j_a} \bar{T}^{(\lambda 0 )\lambda}_{p\bar{q}a\bar{b}} S_{ab} \Biggr)\\
&\quad + \frac{1}{2}\sum_{\substack{abc\\J_1J_2}}
  (n_an_b\bar{n}_c + \bar{n}_a\bar{n}_bn_c)
  \hat{J}_1\hat{J}_2(-1)^{j_p+j_c+J_1+\lambda}
   \begin{Bmatrix}
     J_1 & J_2 & \lambda \\
     j_q & j_p & j_c
   \end{Bmatrix}
    \left( \unnorm{S}^{J_1}_{cpab}\unnorm{T}^{(J_1J_2)\lambda}_{abcq}-\unnorm{T}^{(J_1J_2)\lambda}_{cpab} \unnorm{S}^{J_2}_{abcq} \right)
\end{aligned}
\label{eq:comm1body}
\end{equation}
and
\begin{equation}
\begin{aligned}
\unnorm{C}^{(J_1J_2)\lambda}_{pqrs}
 &= \sum_{a} \left( S_{pa} \unnorm{T}^{(J_1J_2) \lambda}_{aqrs} 
  + S_{qa} \unnorm{T}^{(J_1J_2) \lambda}_{pars}
  - \unnorm{T}^{(J_1J_2) \lambda}_{pqas}  S_{ar}
  -  \unnorm{T}^{(J_1J_2) \lambda}_{pqra}  S_{as} \right) \\
 &\quad -\hat{J}_1\hat{J}_2(-1)^{\lambda} \sum_{a} \Biggl[
     \hspace{0.2cm}(1-P_{pq}(J_1))(-1)^{j_p+j_q+J_2}
    \begin{Bmatrix}
      J_2 & J_1 & \lambda \\
      j_p & j_a & j_q
    \end{Bmatrix}
    T^{\lambda}_{pa} \unnorm{S}^{J_2}_{aqrs} \\
     &\hspace{3.1cm}-(1-P_{rs}(J_2))(-1)^{j_r+j_s-J_1}
    \begin{Bmatrix}
      J_1 & J_2 & \lambda \\
      j_s & j_a & j_r
    \end{Bmatrix}
    \unnorm{S}^{J_1}_{pqra} T^{\lambda}_{as}
 \hspace{0.2cm} \Biggr] \\
 &\quad + \frac{1}{2}\sum_{ab}(1-n_a-n_b)(\unnorm{S}^{J_1}_{pqab}\unnorm{T}^{(J_1J_2)\lambda}_{abrs}- \unnorm{T}^{(J_1J_2)\lambda}_{pqab} \unnorm{S}^{J_2}_{abrs})\\
 &\quad + \sum_{abJ_3J_4}\hat{J}_1\hat{J}_2\hat{J}_3\hat{J}_4
   (n_a-n_b) \Biggl[(1-P_{pq}(J_1))(1-P_{rs}(J_2))
   (-1)^{j_q+j_s+J_2+J_4}
   \begin{Bmatrix}
    j_p & j_s & J_3 \\
    j_q & j_r & J_4 \\
    J_1 & J_2 & \lambda
   \end{Bmatrix}
   \bar{S}^{J_3}_{p\bar{s}a\bar{b}} \bar{T}^{(J_3J_4)\lambda}_{a\bar{b}r\bar{q}}
   \Biggr].
\end{aligned}
\label{eq:comm2body}
\end{equation}
In equations (\ref{eq:comm1body}) and (\ref{eq:comm2body}), $\hat{J}\equiv \sqrt{2J+1}$, $n_a$ is the occupancy of orbit $a$, with $0\leq n_a \leq 1$, and $\bar{n}_a \equiv (1-n_a)$, and $P_{pq}(J) \equiv (-1)^{j_p + j_q -J} P_{pq}$ is the spherical-basis permutation operator.
We have also employed the Pandya-transformed operators defined by
\begin{equation}
\bar{S}^{J_1}_{p\bar{q}r\bar{s}} = - \sum_{J_2}\hat{J}_2
\begin{Bmatrix}
j_p & j_q & J_1 \\
j_r & j_s & J_2
\end{Bmatrix}
\unnorm{S}^{J_2}_{psrq}\,,
\end{equation}
\begin{equation}
\bar{T}^{(J_1J_2)\lambda}_{p\bar{q}r\bar{s}} =
-\sum_{J_3J_4}\hat{J}_1\hat{J}_2\hat{J}_3\hat{J}_4
(-1)^{j_q +j_s+J_2+J_4}
\begin{Bmatrix}
j_p & j_s & J_3 \\
j_q & j_r & J_4 \\
J_1 & J_2 & \lambda
\end{Bmatrix}
\unnorm{T}^{(J_3J_4)\lambda}_{psrq}.
\end{equation}

For computation of electromagnetic moments with EOM-IMSRG or transitions between multiple EOM excited states, expressions for the product of two spherical tensors of arbitrary rank are needed. The tensor product is given by 
\begin{equation}
\mathcal{Y}^J_M \equiv [\mathcal{O^\lambda} \times X^\dagger_\nu(J_\nu)]^J_M = \sum_{M_\nu \mu} C^{\lambda J_\nu J}_{\mu M_\nu M} \mathcal{O}^\lambda_\mu X^\dagger_\nu( J_\nu M_{\nu}).
\label{eq:tensor_prod_def_APP}
\end{equation}
where
\begin{equation}
\label{eq:tensor_prod_1b}
\begin{aligned} 
Y^J_{pq} &= \hat{J}(-1)^{(j_p + j_q)} \sum_a (O^{\lambda}_{pa} X^{{J_\nu}}_{aq}(-1)^J \sj{\lambda}{{J_\nu}}{J}{j_q}{j_p}{j_a} \bar{n}_a
- X^{{J_\nu}}_{pa} O^{\lambda}_{aq}(-1)^{(\lambda + {J_\nu})} \sj{\lambda}{{J_\nu}}{J}{j_p}{j_q}{j_a} n_a ) \\ 
&~+ \sum_{ab} (\frac{1}{\hat{\lambda}} O^{\lambda}_{ba} \bar{X}^{(J \lambda){J_\nu}}_{p \bar{q} b \bar{a}}
+ \frac{(-1)^{(\lambda + {J_\nu} +J)}}{\hat{{J_\nu}}}\bar{O}^{ (J {J_\nu})\lambda}_{p \bar{q} a \bar
{b}} X^{{J_\nu}}_{ab}  )n_b \bar{n}_a  \\
&~- \frac{1}{2} \sum_{abc} \sum_{J_1 J_2 J_3} \hat{J}\hat{J_1}\hat{J_3}\sj{j_p}{j_q}{J}{J_3}{J_1}{j_c} \sj{\lambda}{{J_\nu}}{J}{J_3}{J_1}{J_2} \unnorm{O}^{ (J_1 J_2)\lambda}_{cpab}  \unnorm{X}^{ (J_2 J_3){J_\nu}}_{abqc} \bar{n}_a \bar{n}_b n_c\\
&~+ \frac{1}{2} (-1)^{(\lambda+{J_\nu} +J)}\sum_{abc} \sum_{J_1 J_2 J_3}\hat{J}\hat{J_1}\hat{J_3}  \sj{j_p}{j_q}{J}{J_3}{J_1}{j_c} \sj{J_\nu}{\lambda}{J}{J_3}{J_1}{J_2} \unnorm{X}^{ (J_1 J_2){J_\nu}}_{cpab}  \unnorm{O}^{ (J_2 J_3)\lambda}_{abqc} n_a n_b \bar{n}_c
\end{aligned}
\end{equation}
and
\begin{equation}
\label{eq:tensor_prod_2b}
\begin{aligned} 
\unnorm{Y}^{(J_1 J_2) \lambda}_{pqrs} = (1-P_{pq}(J_1))& (1-P_{rs}(J_2)) \hat{\lambda} \hat{\lambda}_1 \hat{\lambda}_2 \nj{j_p}{j_r}{\lambda_1}{j_q}{j_s}{\lambda_2}{J_1}{J_2}{\lambda} O^{\lambda_1}_{pr} \unnorm{X}^{\lambda_2}_{qs} \\ 
+(1-P_{pq}(J_1)) \hat{\lambda}\hat{J_1}\sum_a \sum_{J_3}\hat{J_3}&(-1)^{(j_p + j_q + J_1 +J_2 +J_3 + \lambda_1 + \lambda)} \sj{j_p}{j_q}{J_1}{J_3}{\lambda_1}{j_a}
\sj{\lambda_1}{\lambda_2}{\lambda}{J_2}{J_1}{J_3} O^{\lambda_1}_{pa} \unnorm{X}^{ (J_3 J_2)\lambda_2}_{aqrs} \bar{n}_a \\ 
+ (1-P_{rs}(J_2))\hat{\lambda}\hat{J_2}\sum_a \sum_{J_3} \hat{J_3} &(-1)^{(J_1+J_3+\lambda_2)} \sj{j_r}{j_s}{J_2}{J_3}{\lambda_1}{j_a}\sj{\lambda_1}{\lambda_2}{\lambda}{J_1}{J_2}{J_3}
O^{\lambda_1}_{ar} \unnorm{X}^{ (J_1 J_3)\lambda_2}_{pqsa} n_a \\ 
- (1-P_{rs}(J_2))\hat{\lambda}\hat{J_2}\sum_a \sum_{J_3} \hat{J_3} &(-1)^{(J_1+J_3+\lambda_2 + \lambda)} \sj{j_r}{j_s}{J_2}{J_3}{\lambda_2}{j_a}
\sj{\lambda_1}{\lambda_2}{\lambda}{J_2}{J_1}{J_3} X^{\lambda_2}_{ar} \unnorm{O}^{ (J_1 J_3)\lambda_1}_{pqsa} \bar{n}_a\\
- (1-P_{pq}(J_1)) \hat{\lambda}\hat{J_1} \sum_a \sum_{J_3} \hat{J_3} &(-1)^{(j_p + j_q + J_1 +J_2 +J_3 + \lambda_1)} \sj{j_p}{j_q}{J_1}{J_3}{\lambda_2}{j_a}
\sj{\lambda_1}{\lambda_2}{\lambda}{J_1}{J_2}{J_3} X^{\lambda_2}_{pa} \unnorm{O}^{ (J_3 J_2)\lambda_2}_{aqrs} n_a\\ 
+ \frac{1}{2} \hat{\lambda} \sum_{ab} \sum_{J_3} &(-1)^{(J_1+J_2 + \lambda)} \sj{\lambda_1}{\lambda_2}{\lambda}{J_2}{J_1}{J_3} \unnorm{O}^{ (J_1 J_3)\lambda_1}_{pqab}
\unnorm{X}^{ (J_3 J_2)\lambda_2}_{abrs} \bar{n}_a \bar{n}_b\\ 
+ \frac{1}{2} \hat{\lambda} \sum_{ab} \sum_{J_3} &(-1)^{(J_1+J_2 + \lambda_1 + \lambda_2)} \sj{\lambda_1}{\lambda_2}{\lambda}{J_1}{J_2}{J_3} \unnorm{X}^{ (J_1 J_3)\lambda_2}_{pqab}
\unnorm{O}^{ (J_3 J_2)\lambda_1}_{abrs} n_a n_b\\
+ (1 - P_{pq}(J_1)) &(1-P_{rs}(J_2)) \hat{\lambda}\hat{J_1}\hat{J_2}\sum_{J_3 J_4 J_5}\hat{J_{3}}\hat{J_5} (-1)^{(j_s-j_q + J_3 + \lambda)}\\
\times \nj{j_p}{j_r}{J_3}{j_q}{j_s}{J_5}{J_1}{J_2}{\lambda} &\sj{\lambda_1}{\lambda_2}{\lambda}{J_5}{J_3}{J_4} \sum_{ab} \bar{O}^{ (J_3 J_4)\lambda_1}_{p \bar{r} a \bar{b}}
\bar{X}^{ (J_4 J_5)\lambda_2}_{a \bar{b} s \bar{q}} \bar{n}_a n_b 
\end{aligned}
\end{equation}

\section{Angular momentum coupling identities}
The following identities are helpful in deriving the equations in Appendix~\ref{sec:APP_products}.
\begin{equation}\label{eq:CGID1}
\begin{aligned}
\sum_{m_1 M_1 M_2} &\cg{j_1}{j_2}{J_1}{m_1}{m_2}{M_1}
\cg{j_1}{j_3}{J_2}{m_1}{m_3}{M_2}
\cg{J_2}{J_3}{J_1}{M_2}{M_3}{M_1}
=\frac{\hat{J}_1^{2}\hat{J}_2}{\hat{j}_1}
(-1)^{j_1+j_3+J_1+J_3}
\sj{J_1}{J_2}{J_3}{j_3}{j_2}{j_1}
\cg{j_3}{J_3}{j_2}{m_3}{M_3}{m_2}
\end{aligned}
\end{equation}
\begin{equation}
\begin{aligned}
\sum_{m_1m_2M_1} &
\cg{j_1}{j_2}{J_1}{m_1}{m_2}{M_1}
\cg{j_3}{j_4}{J_1}{m_3}{m_4}{M_1}
\cg{j_1}{J_2}{j_3}{m_1}{M_2}{m_3}
=\frac{\hat{J}_1^2\hat{j}_3}{\hat{j}_2}
(-1)^{j_1+j_2+J_1}
\sj{j_3}{j_1}{J_2}{j_2}{j_4}{J_1}
\cg{j_4}{J_2}{j_1}{m_4}{M_2}{m_2}
\end{aligned}
\end{equation}
\begin{equation}
\begin{aligned}
\sum_{m_1}&
\cg{j_1}{J_1}{j_2}{m_1}{M_1}{m_2}
\cg{j_1}{j_3}{J_2}{m_1}{m_3}{M_2} 
=
(-1)^{j_2+j_3+J_1+J_2} 
\times
\sum_{J_3M_3}\hat{J}_2\hat{j}_{2}
\sj{J_2}{J_3}{J_1}{j_2}{j_1}{j_3}
\cg{J_2}{J_1}{J_3}{M_2}{M_1}{M_3}
\cg{j_2}{j_3}{J_3}{m_2}{m_3}{M_3}
\end{aligned}
\end{equation}
\begin{equation}
\begin{aligned}
\sum_{M_1}&
\cg{j_1}{j_2}{J_1}{m_1}{-m_2}{M_1}
\cg{j_3}{j_4}{J_1}{m_3}{-m_4}{M_1}
= 
\sum_{J_2M_2} \hat{J}^2_2
\sj{j_1}{j_4}{J_2}{j_3}{j_2}{J_1}
\cg{j_1}{j_4}{J_2}{m_1}{m_4}{M_2}
\cg{j_3}{j_2}{J_1}{m_3}{m_2}{M_1}
\end{aligned}
\end{equation}
\begin{equation}
\begin{aligned}
\sum_{M_1M_2} &
\cg{j_1}{j_2}{J_1}{m_1}{-m_2}{M_1}
\cg{j_3}{j_4}{J_2}{m_3}{-m_4}{M_2}
\cg{J_2}{J_3}{J_1}{M_2}{M_3}{M_1}
=\sum_{\substack{J_4 J_5 \\ M_4 M_5}}
\hat{J}_1^2\hat{J}_2\hat{J}_4
(-1)^{j_2+j_4+J_1+J_4}
\nj{j_1}{j_2}{J_1}{j_4}{j_3}{J_2}{J_4}{J_5}{J_3} 
\times
\cg{j_1}{j_4}{J_4}{m_1}{m_4}{M_4}
\cg{j_3}{j_2}{J_5}{m_3}{m_2}{M_5}
\cg{J_5}{J_3}{J_4}{M_5}{M_3}{M_4}
\end{aligned}
\end{equation}
\end{widetext}

\end{document}